%% file: ms.tex
\documentclass[3p,review]{elsarticle}
\usepackage[utf8]{inputenc}
\usepackage{amsmath,amssymb,amsfonts,latexsym,cancel}
\usepackage{graphicx,caption,subfigure,float,esvect}
\usepackage{accents}
\usepackage{booktabs}
\usepackage[rflt]{floatflt}
\usepackage{xcolor}
\newcommand\mnewcommand[1]{% 
\let#1\relax \newcommand#1 }
\mnewcommand{\isdisc}[1]{\boldsymbol{#1}}
\mnewcommand{\isvector}[1]{\boldsymbol{#1}}
\mnewcommand{\isFtr}[1]{\widehat{#1}}
\mnewcommand{\isApp}[1]{\widetilde{#1}}
\mnewcommand{\isoperator}[1]{\mathsf{#1}}
\mnewcommand{\istr}[1]{\widehat{#1}}
\mnewcommand{\innprod}[3]{\left \langle #1 \,|\, #2 \right \rangle_{#3}}
\mnewcommand{\isconj}[1]{\overline{#1}}
\mnewcommand{\Reals}{\mathrm {I\!R}}
\mnewcommand{\Imags}{\mathrm {I\!I}}
\mnewcommand{\Comxs}{\mathbb{C}}

\graphicspath{{}}

\begin{document}

\begin{frontmatter}

\title{A general method to compute numerical dispersion errors and its application to stretched meshes}

\author[cttc]{J.~Ruano\corref{cor1}}
\ead{jesusr@cttc.upc.edu}

\author[cttc]{A.~Baez Vidal\corref{cor1}}
\ead{aleix@cttc.upc.edu}

\author[cttc]{F.X.~Trias}
\ead{xavi@cttc.upc.edu}

\author[cttc]{J.~Rigola}
\ead{quim@cttc.upc.edu}

\cortext[cor1]{Corresponding author}
\address[cttc]{Heat and Mass Transfer Technological Center (CTTC), 
Universitat Politècnica de Catalunya – BarcelonaTech
(UPC) ETSEIAT, Colom 11, 08222, Terrassa, Barcelona, Spain}

\begin{abstract}
This article presents a new spectral analysis approach for dispersion error and a
methodology to numerically evaluate it. In practice, this new analysis allows the numerical 
study of dispersion errors on all types of mesh and for multiple dimensions. 
Nonetheless, when mesh uniformity and one-dimensionality
assumptions are imposed as in the classical method, the results of this
new technique coincide with those of the classic method. We establish
the theoretical basis of the approach, derive a numerical methodology to
evaluate dispersion errors and assess the method after a set of numerical
tests on non-uniform stretched meshes.
\end{abstract}


\end{frontmatter}

%\maketitle

\section{Introduction}
\input{1.Introduction.tex}

\section{Analytical derivation}
\input{2.AnalyticalDerivation.tex}

\section{Methodology}
\input{3.MethodologyGeneral.tex}
\subsection{Evaluating the gradient}
\input{4.GradientEval.tex}
\subsection{Selection of an orthonormal basis}
\input{5.OrthonormalBasis.tex}

\subsection{Studying sinusoids or other functions}
\input{6.ChangeBasis.tex}

\section{Numerical tests}
\input{7.ApplicationCase.tex}

\section{Conclusions}
\input{8.Conclusions.tex}

\section*{Acknowledgements}
This work has been financially supported by the \textit{Ministerio de
Econom\'ia y Competitividad}, Spain (No. ENE2017-88697-R). 
J.R.P. is supported by a \textit{FI-DGR 2015} predoctoral contract financed by 
\textit{Generalitat de Catalunya}, Spain. 
F.X.T.  is supported by a \textit{Ram\'on y Cajal} postdoctoral Contract (No.  RYC-2012-11996) 
financed by the \textit{Ministerio de Econom\'ia y Competitividad}, Spain. 

\section*{References}
\bibliographystyle{abbrv}
\bibliography{biblio}

\end{document}

%% file: 1.Introduction.tex
Computational Fluid Dynamics (CFD) is the branch of physics that computes and analyses numerical
simulations that involve fluid flow. The branch of CFD that studies the
generation and propagation of sound caused by fluid flow is known as
Computational Aeroacoustics (CAA). In CAA, it is essential to keep
accuracy to simulate the propagating speed of sound waves. Else, the
interference patterns are erroneous and the simulations are irrelevant. 
The main scheme-depending factor affecting the simulated wave propagation velocity is the specific
discretization of the differential operator of the equations ruling the 
fluid movement. In this regard, the classical dispersion error analysis
\cite{Lele1992} concludes that high order numerical schemes alleviate the committed propagation error. 

The classical methodology applies on the transport term of the system of Partial
Differential Equations modelling wave propagation. However, it requires this term
to be in advective form. For example, for the Euler Equations, one should use
their characteristics form, i.e., the
fluxes are calculated after applying the chain rule for
the derivatives 
\begin{equation} \label{eq:EulerEqs}
\frac{\partial}{\partial t} \isdisc{q} +
\nabla\cdot\isdisc{f}(\isdisc{q})=\frac{\partial}{\partial t} \isdisc{q} +
J(\isvector{f},\,\isvector{q})\cdot\nabla\isdisc{q}=\isdisc{0}; 
\end{equation}
where $\isdisc{q}\in\Reals^{D+2}$ are the fluid magnitudes,
$\isdisc{f}\in\Reals^{D+2}$ Eulerian fluxes, 
$J(\isvector{f},\,\isvector{q})\in\Reals^{(D+2)\times(D+2)}$ the Jacobian
$J_{ij}=\partial{f_i}/\partial{q_j}$ and $D$ the number of dimensions
of the model. Then, the term in the center of
Eq.(\ref{eq:EulerEqs}) is projected on a unitary orthonormal basis of
the Euclidean space where it is defined and one obtains the set of locally
decoupled wave equations on each of the 
characteristic variables
\begin{equation}\label{eq:WaveEq}
\frac{\partial }{\partial t} r_j + c_j \frac{\partial}{\partial s} r_j=0,\;\qquad j\in[1,\,D+2],
\end{equation}

where $t$ is the time and $s$ is a generic space coordinate. Then, the analysis studies the 
ratio between numerical approximations of $\partial r_j/\partial s$ and $r_j$ in Eq.(\ref{eq:WaveEq}) 
when $r_j$ are mono-modal sinusoids and the analytical values these ratios should take. 
These ratios are evaluated in the Fourier space by means of the Fourier Transform algorithm. 
The wave propagation speeds $c_j$ in this procedure are assumed to be constant in space and time.

Tam et al. \cite{Tam2012,Tam1993} used the classical
analysis to develop numerical methods that reduce such errors by following the relationship
between wavenumber and wave speed propagation, i.e. the dispersion
relation.  
The extracted conclusions of both analysis, summarized in Colonius \cite{Colonius2004},
have favored the use of high order methods in wave propagation problems.
For example, Bogey et al. \cite{Bogey2018} used high order schemes to aleviate
the dispersion error \cite{Bogey2004} to simulate flow and acoustic fields of high-Reynolds jets.
Shur et al. \cite{Shur2005,Shur2016} also used a fourth order central - fifth order upwind variant
of the Roe scheme to compute the inviscid fluxes in their jets simulation. 
Bodony \cite{Bodony2008} compiled a list of several works in the field of jet-noise 
computation which used high order approximations of the convective term.
Whereas previous authors solved Navier-Stokes equations to compute noise, Ewert \cite{Ewert2003} 
and Seo \cite{Seo2006} solved additional differential equations. Nevertheless, they also used
high order approximations of the transport term to reduce the dispersion. 

However, the range of validity of the classical
dispersion error analysis is limited due to the assumptions it lies on:
\begin{itemize}
 \item Uniform mesh/spacing: The function at neighbour nodes, control volumes or points, 
%is expressed as the function at  the central node of the stencil by a shift. 
 is expressed as a shift of the function at the central node of the stencil.
 \item Linear wave propagation: The convective operator is linearized
 into the gradient operator multiplied by a constant propagation speed
($c_j$ in Eq.(\ref{eq:WaveEq}) is constant).
 \item Linear schemes: The coefficients of the stencil are constant during
 the simulations and not dependent on the obtained fields, i.e., a linear discrete operator.
\end{itemize}
Consequently, the theoretical framework of
\cite{Lele1992,Tam2012,Tam1993} does not cover all the
possible situations in CAA. In addition, no means to extend the validity
of the analysis are given. 

The limitations of the classical analysis become apparent when
simulating flows in 2 or 3 space dimensions. Then, a common approach is to apply one stencil 
for each direction and assume
that the overall method reduces the overall dispersion error
because each direction stencil does. However, wave propagation in 2D and
3D brings issues like anisotropy and wavenumber directionality that cannot 
be modelled in 1D and are forfeit when choosing or designing
numerical schemes.

Another limitation of the classical analysis comes when trying to take
into account the effect of non-linearity of convection on dispersion
error. To do it, Pirozzoli \cite{Pirozzoli2006}
developed a different methodology based on a numerical single time step integration. 

Finally, the classical analysis does not allow to study mesh non-uniformity 
in regard of the  properties of the numerical schemes. Such analyses
would have a great impact since most of the meshes for CAA simulations are
stretched in the direction of wave propagation; see Bogey \cite{Bogey2018}, 
Shur \cite{Shur2005,Shur2016} or  Bodony \cite{Bodony2008}.
Even in the case when the extracted conclusions were mesh dependant, a dispersion
error analysis that accepted non-uniform discretizations would help to
understand the behaviour of the differencing scheme on non-uniform
meshes and designing meshes accordingly. 

Only Trefethen \cite{Trefethen1982} and Vichnevetsky \cite{Vichnevetsky1981}
performed this kind of analysis examining how grid stretching and anisotropy, 
or non-constant coefficients affected the accuracy of wave propagation and generation
of dispersive waves.
The Fourier Transform of a function is the projection on the space of the eigenfunctions
of the countinuous Laplacian operator \cite{Fourier1822}. Once the space is discretized, the 
discrete equivalent of the continuous Laplacian is a discrete Laplacian, and the equivalent
of its eigenfunctions are now eigenvectors. Both sets of eigenfunctions and eigenvectors
are orthogonal. When evaluating the eigenfunctions of the continuous Laplacian on a discretization,
the resulting set of vectors are not orthogonal unless the discretization is uniform. The Discrete
Fourier Transform (DFT) lies on this assumption. Henceforth, the DFT is not well fit for analysing
dispersion errors on non evenly-spaced discretizations.

To sum up, the main reason for using high order methods to reduce dispersion error in wave
propagation simulations seems to be weak when studying flow-depending
operators, multiple dimensions or non-uniform meshes. Here we develop a new methodology to overcome the limitations of the classical approach. 
This opens the door to dispersion error analysis in any possible mesh. 
The presented methodology is tested in a Finite Volume Method (FVM) framework. 

The rest of the paper is organized as follows: in section 2 the link between the classical approach
used to study dispersion error and the present methodology is discussed as well as all the analytical
derivation required. In section 3 we detail how to apply this alternative methodology. In section 4, a set of 
numerical tests are proposed and conducted. The results are latter analysed and commented. 
Finally, in section 5 the extracted conclusions of the whole paper are discussed.

%% file: 2.AnalyticalDerivation.tex
\label{sec:AnalDerivation}

Let $f(x):\Reals\mapsto\Reals$ be a function that could 
be decomposed into a sum of sinusoids and 
\begin{equation}
% \isFtr{f}=\displaystyle{{\int_{\Omega} f(x)\cdot e^{-2{\pi}ix\alpha}\,dx}}
 \isFtr{f}=\displaystyle{{\int\limits_{-\infty}^{\infty} f(x) e^{-2{\pi}ix\alpha}\,dx}},
\end{equation}
its Fourier Transform. The Fourier Transform of the derivative 
\begin{equation}
f'=\frac{d}{dx}f(x), 
\end{equation}
of $f$ is:
\begin{equation}\label{dispRel}
\isFtr{f}'(\alpha)=i\alpha\isFtr{f}.
\end{equation}

However, the previous expressions do not hold when the space is discretized.
Taking a discretization of the physical space defined by
$\Omega=\{\omega_1(x),\, \omega_2(x),\, \omega_3(x),\,\ldots,\omega_N(x)\}^T$, $f$ is
approximated with
\begin{equation}\label{eq:discdef}
f(x)\simeq \sum_{j=1}^{N} \omega_j(x) f_j=\Omega\cdot\isdisc{f},
\end{equation}
where the "$N$" scalars $f_j$ are the components of the array
$\isdisc{f}=\{f_1,\,f_2,\,f_3,\,\ldots\,,f_N\}^T\in\Reals^N$. 
If a Finite Difference discretization is selected, the different elements of $\Omega$ would be equal to:
\begin{equation}
 \omega_i(x) = 
 \left\{ 
   \begin{array}{rcl}
1 & \mbox{if} & x=x_i\\
&
& \\
0 & \mbox{elsewhere}
& \\
   \end{array}
 \right.
\end{equation}
In case a Finite Volume approach is selected, with non-overlapping volumes, the 
elements of $\Omega$ would be:
\begin{equation}
 \omega_i(x) = 
 \left\{ 
   \begin{array}{rcl}
1 & \mbox{if} & x\in V_i\\
&
& \\
0 & \mbox{elsewhere}
& \\
   \end{array}
 \right.
\end{equation}
Where $V_i$ denotes the $i^{th}$ control volume of the discretization.
Assuming the discretization is uniform and the physical space is 1D,
one can approximate the derivative $f'$ with 
\begin{equation}\label{eq:expDer}
% {f'}(x) \simeq \sum_j f_j \omega'_j(x)= \sum_j \displaystyle{\left(f_j \frac{1}{\Delta x} \sum_{k=-p}^{p} a_{k}  \omega_{j+k}(x)\right)}, 
% {f'}(x) \simeq \sum_j \displaystyle{\left(f_j \sum_{k=1}^{N} a_{j,k}  \omega_{k}(x)\right)}, 
 {f'}(x) \simeq \sum_{j=1}^{N} \omega_j(x) \sum_{k=1}^{N} a_{j,k} f_k, 
\end{equation}
The previous equation can be written in a matrix-vector form as:
\begin{equation}
% f'(x) = A\cdot f \cdot \omega^T(x),
 f'(x) \simeq {\Omega}^T A\isdisc{f},
\end{equation}
where the matrix $A$ stands for the discrete differential operator. 

Applying the shifting theorem and the derivative theorem (Tam
\cite{Tam2012,Tam1993}), the Fourier Transforms of $f$ and the
discrete approximation to its derivative are related with
\begin{equation}
\isFtr{f}'(\alpha)\approx \displaystyle {\frac{1}{\Delta
x}}\left[\sum_{k}{a_{k}e^{ik\alpha\Delta x}}\right]\isFtr{f}(\alpha).
\label{eq:ApproxDerFourierSpace}
\end{equation}

And the numerical wavenumber $\isApp{\alpha}$ according to the
classical analysis follows straighforwardly:
\begin{equation}
% \begin{array}{c}
%  i\isApp{\alpha}\isFtr{f}=\displaystyle {\frac{1}{\Delta x}}\left[\sum_{k}{a_{k}e^{ik\alpha\Delta x}}\right]\isFtr{f};\\
% \isApp{\alpha}=\displaystyle {\frac{-i}{\Delta x}}\left[\sum_{k}{a_{k}e^{ik\alpha\Delta x}}\right].
% \end{array}
i\isApp{\alpha}\isFtr{f}=\displaystyle {\frac{1}{\Delta x}}\left[\sum_{k}{a_{k}e^{ik\alpha\Delta x}}\right]\isFtr{f}
\Longrightarrow
\isApp{\alpha}=\displaystyle {\frac{-i}{\Delta x}}\left[\sum_{k}{a_{k}e^{ik\alpha\Delta x}}\right].
\end{equation}

But this definition assumes a uniform 1D mesh. To broaden the
concept of numerical wavenumber to non-uniform meshes where
Eq.(\ref{eq:ApproxDerFourierSpace}) does not apply, the eigenvalues
of the derivative operator can be studied. More precisely, the differences 
between the analytical and numerical eigenvalues of the
derivative operator. To extend the concept to non-uniform
discretizations, we realize that the Fourier Transform projects
functions into the space of eigenfunctions of the derivative operator of
evenly distributed domains, i.e., Euclidean spaces or uniformly
discretized domains. Thus, instead of projecting the function $f$ in the
space of sinusoids, we propose to project $f$ into the space of eigenfunctions of the derivative
operator, which does not coincide with sinusoids in non-uniform
discretizations. Then, Eq.(\ref{dispRel}) must be rewritten as:
\begin{equation}
 {f'}^{\dagger}(\lambda)=\lambda{f}^{\dagger}(\lambda),
\end{equation}
where the projection, ${()}^{\dagger}$, is computed as:
\begin{equation} 
 \begin{array}{c}
%  {f'}^{\dagger}=\displaystyle{{\int_{\Omega} f'(x)\cdot\omega dx}}\\
%  {f}^{\dagger}=\displaystyle{{\int_{\Omega} f(x)\cdot\omega dx}}
  {f'}^{\dagger}(\lambda)=\displaystyle{{\int\limits_{-\infty}^{\infty} f'(x)\cdot\beta(\lambda,x) \,dx}};\\
  {f}^{\dagger}(\lambda)=\displaystyle{{\int\limits_{-\infty}^{\infty} f(x)\cdot\beta(\lambda,x) \,dx}},
 \end{array}
\end{equation}
and $\beta$ and $\lambda$ are the appropiate set of eigenfunctions and eigenvalues extracted from the 
first-order derivative operator. If we let
\begin{equation} \label{eq:innProdDef}
\innprod{\xi}{\psi}{\Omega_\mu}={\int\limits_{\Omega_{\mu}} \xi(\mu)
\isconj{\psi}(\mu) \,d\mu}.
\end{equation}
be the inner product of
$\{\xi,\,\psi\}\in \mathcal{L}^{2}(\Omega_\mu,\,\mu)$, where
$\mathcal{L}^{2}(\Omega_\mu, \mu)$ is the space of square
Lebesgue-integrable functions $\psi:\Omega_\mu\subset\Reals\mapsto\Comxs$ and $\isconj{\psi}$ 
the complex conjugate of $\psi$. We can rewrite the projections in a more compact form:
\begin{equation} 
 \begin{array}{c}
%  {f'}^{\dagger}=\displaystyle{{\int_{\Omega} f'(x)\cdot\omega dx}}\\
%  {f}^{\dagger}=\displaystyle{{\int_{\Omega} f(x)\cdot\omega dx}}
  {f'}^{\dagger}(\lambda)=\innprod{f'}{\beta}{\Omega_x};\\
  {f}^{\dagger}(\lambda)=\innprod{f}{\beta}{\Omega_x}.
 \end{array}
\end{equation}

Once discretized, an implicit relation between the numerical eigenvalue and the
analytical one is found:
\begin{equation}\label{eigenLambda}
\isApp{\lambda}=\displaystyle{\frac{\isApp{{f'}^{\dagger}}}{\isApp{{f}^{\dagger}}}}=
\displaystyle{\frac{\innprod{A\isdisc{f}}{\isdisc{\beta}}{}}{\innprod{\isdisc{f}}{\isdisc{\beta}}{}}},
\end{equation}
where the discrete inner product is defined as:
\begin{equation}
\innprod{\isdisc{\phi}}{\isdisc{\psi}}{}=
\sum_{j=1}^N \phi_j
\sum_{k=1}^N \innprod{\omega_j}{\omega_k}{\Omega_x}\isconj{\psi}_k.
\end{equation}
We remark that $\omega_j$ and $\omega_k$ are the $j$ and $k$ terms of the discretization $\Omega$
defined at the beginning of this section.

Both methodologies, the classical and that defined in
Eq.(\ref{eigenLambda}), assume that the derivative can be explicitly projected into a selected
space of functions and depends linearly on the original $f$. For more
complicated derivation processes, e.g. non-linear differential
operators, these expressions are not valid. Consequently, another method
is proposed.

Namely, let $\Phi=\{\phi_{-N}(x),
\phi_{-N+1}(x),\,\ldots\phi_{-1}(x),\,\phi_0(x),\,\phi_1(x),\,\phi_2(x),\,\ldots\phi_N(x)\}$ be an
orthonormal basis of functions of $\Omega_x\subset\Reals$, i.e, 
\begin{equation}
\innprod{\phi_m}{\phi_n}{\Omega_x} = \delta_{mn},
\end{equation}
where $\delta_{mn}$ is the Kronecker's delta.

One can thus define a mapping
$T:\mathcal{L}^{2}(\Omega_x, x)\mapsto \Comxs^{2N+1}$;
% $T:f(x)\mapsto (\alpha_m)_{m\in\Comxs}$, where 
 $T:f(x)\mapsto (\alpha_m)\in\Comxs^{2N+1}$, where 

\begin{equation}
\alpha_m=\innprod{f}{\phi_m}{\Omega_x}={\int\limits_{\Omega_x}
f(x)\isconj{\phi_m}(x)  \,dx}, 
\end{equation}
is the projection of $f(x)$ onto the $m$ function $\Phi$.
Under the pertinent smoothness of $f$ criterion,
\begin{equation}
f(x)\simeq S_N=\sum_{m=-N}^{N}\alpha_m\phi_m(x); \quad
\lim_{N\rightarrow\infty} S_N=f(x).
\end{equation}
This defines the inverse mapping
$T^{-1}:\Comxs^{2N+1}\mapsto\mathcal{L}^{2}(\Omega_x, x)$,
%$T:(\alpha_m)_{m\in\Comxs}\mapsto f(x)$.
$T:(\alpha_m)\in\Comxs^{2N+1}\mapsto f(x)$.

The derivative $f'=\frac{d}{dx}f(x)$ can also be expressed in terms of its
projection on the set of functions of $\Phi$: 
\begin{equation}\label{diff1}
%% f'(x)=\displaystyle{{\int_{\Gamma} \beta(m)\cdot\phi(x,m) dm}}
 f'(x)\simeq S'_N= \sum_{m=-N}^{N}\alpha_m\phi'_m(x)\simeq
\sum_{m=-N}^{N}\alpha_m \sum_{n=-N}^{N} \gamma_{mn}\,\phi_n(x),
\end{equation}
where 
\begin{equation}\label{definitionGamma}
\gamma_{mn}=\innprod{\phi'_m}{\phi_n}{\Omega_x}.
\end{equation}

%Both $\alpha$ and $\beta$ coefficients are linked to the selected function
%$f(x)$ and the selected basis of functions. But 
%$f'$ can
%also be expressed as 
%\begin{equation}
%% f'(x)=\frac{d}{dx}\displaystyle{{\int_{\Gamma} \alpha(m)\cdot\phi(x,m) dm}}=
%%       \displaystyle{{\int_{\Gamma} \alpha(m)\cdot\frac{\partial\phi(x,m)}{\partial x} dm}}=
%%      \displaystyle{{\int_{\Gamma} \alpha(m)\cdot\phi'(x,m) dm}}
% f'(x)=\frac{d}{dx}\innprod{\alpha}{\Phi(x,\,m)}{m}=
%\innprod{\alpha}{\frac{\partial\phi(x,m)}{\partial
%x}}{m}=\innprod{\alpha}{\Phi_x(x,\,m)}{m},
%\end{equation}
This holds on the orthonormality of the
functions of $\Phi$:

\begin{equation}
\gamma_{mn}
=\innprod{\phi'_m}{\phi_n }{\Omega_x}
=\innprod{\sum_{p=-N}^{N}\gamma_{mp}\,\phi_p(x)}{\phi_n}{\Omega_x}           
=\sum_{p=-N}^{N}\gamma_{mp}\innprod{\phi_p}{\phi_n}{\Omega_x}
=\sum_{p=-N}^{N}\gamma_{mp}\delta_{pn}=\gamma_{mn}.
\end{equation}

%\begin{equation}
%\begin{array}{ll}
%\gamma_{mn}&=\innprod{\phi'_m}{\phi_n }{\Omega_x}\\
%&\displaystyle{=\innprod{\sum_{p=-N}^{N}
%\gamma_{mp}\,\phi_p(x)}{\phi_n}{\Omega_x}}\\           
%&\displaystyle{=\sum_{p=-N}^{N}\gamma_{mp}\innprod{\phi_p}{\phi_n}{\Omega_x}}\\
%&\displaystyle{=\sum_{p=-N}^{N}\gamma_{mp}\delta_{pn}=\gamma_{mn}}.
%\end{array}
%\end{equation}

%This holds on Fubini's theorem and on choosing $\Phi(x,\,n)$ such that
%its iso-$m$ parametric curves are orthonormal:
%\begin{equation}
% \begin{array}{ll}
%%  \gamma(m,n) & =\displaystyle{{\int_{\Omega} \left(\displaystyle{{\int_{\Gamma} \gamma(m,p)\phi(p,x) dp}}\right)\phi(n,x) dx}};\\
%%  & =\displaystyle{{\int_{\Gamma} \left(\displaystyle{{\int_{\Omega} \phi(n,x)\phi(p,x)dx}}\right)\gamma(m,p)dp}};\\
%%  & =\displaystyle{{\int_{\Gamma} \gamma(m,n) dp}};\\
%%  & =\gamma(m,n)\displaystyle{{\int_{\Gamma} dp}};\\
%%  & =\gamma(m,n)
%  \gamma(m,n) & =\displaystyle{{\int\limits_{-\infty}^{\infty} \left(\displaystyle{{\int\limits_{-\infty}^{\infty} \gamma(m,p)\phi(p,x) dp}}\right)\phi(n,x) dx}};\\
%   & =\displaystyle{{\int\limits_{-\infty}^{\infty} \left(\displaystyle{{\int\limits_{-\infty}^{\infty} \phi(n,x)\phi(p,x)dx}}\right)\gamma(m,p)dp}};\\
%   & =\displaystyle{{\int\limits_{-\infty}^{\infty} \gamma(m,n) dp}};\\
%   & =\gamma(m,n)\displaystyle{{\int\limits_{-\infty}^{\infty} dp}};\\
%   & =\gamma(m,n).
% \end{array}
%\end{equation}

With these definitions, $\gamma_{mn}$ are the elements of a matrix
$\Gamma\in\Comxs^{2N+1\times 2N+1}$ and the projections of the
derivatives of $\phi_m$ with respect to ``$x$'' on $\phi_n$. 
%Only the selected basis of functions $\Phi$ and the properties of the
%differential operator under study (the derivative) influence $\Gamma$. 
$\Gamma$ characteristics are determined by the selected basis of 
functions $\Phi$ and the properties of the derivative operator. 
Assuming null contributions from boundaries and integrating by parts, it is straightforward to show 
that the derivative operator is skew-Hermitian with respect to the inner
product of Eq.(\ref{eq:innProdDef}), i.e., 
\begin{equation}
\innprod{\frac{d}{dx}\xi}{\psi}{\Omega_x}=-\innprod{\xi}{\frac{d}{dx}\psi}{\Omega_x}. 
\end{equation}
Thus, $\Gamma$ should be skew-Hermitian too.

Using an approximate derivative operator $\isApp{\frac{d}{dx}}(\cdot)$, one gets
$\isApp{\Gamma}\in\Comxs^{2N+1\times 2N+1}$ instead. Making a parallelism
with Eq.(\ref{eigenLambda}), the errors associated with using
approximate differential operators to approximate derivatives are the
deviations of $\isApp{\gamma}_{mn}$ from $\gamma_{mn}$.  Finally, to ease the
analysis, $\Phi$ can be chosen such that  $\Gamma$ is diagonal
and known, i.e., $\phi_m$ is an eigenfunction of the derivative. Then,
the possible errors in $\isApp{\Gamma}$ are:
\begin{itemize}
 \item {$\isApp{\gamma}_{mn}\neq0$ if $m\neq n$,}
 \item {$Re(\isApp{\gamma}_{mm})\neq0$, and }
 \item {$\displaystyle{\frac{Im(\isApp{\gamma}_{mm})}{\lambda_m}}\neq1$,}
\end{itemize}
where $\lambda_m\in\Imags$ is the eigenvalue of the derivative on $\phi_m$. A methodology to 
compute these errors is described in the next section.

%% file: 3.MethodologyGeneral.tex
\label{sec:Method}
When functions are approximated with discretizations as in Eq.(\ref{eq:discdef}) 
to operate with arrays of scalars, the differential
operators of the equations describing some physical phenomena should
be approximated accordingly. For example, if $\isdisc{f}$ is the
discrete representation of $f(x)$ on $\Omega=\{\omega_1(x),\,
\omega_2(x),\, \omega_3(x),\,\ldots,\omega_N(x)\}$ as defined in Eq.(\ref{eq:discdef}),
the discrete representation of an approximation to its derivative
$\isdisc{f'}$ is represented on 
%$\Omega'= \{\omega'_1(x),\, \omega'_2(x),\, \omega'_3(x),\,\ldots,\omega'_M(x)\}$. 
$W= \{w_1(x),\, w_2(x),\, w_3(x),\,\ldots,w_M(x)\}$. 
The different discretizations methodologies of CFD contemplate $W\neq\Omega$.
However it is common to project the
calculated derivatives onto the original $\Omega$ (see, e.g.
\cite{Trias2014a}) in following computation
steps. Here, we focus on the
compound process $A:\Reals^N\mapsto\Reals^N$, i.e. the
approximation to the derivative
$\isdisc{f}\mapsto A(\isdisc{f})$ and its projection onto $\Omega$. Since differential
operators are linear, their discrete counterparts should be linear too.
Therefore, $A(\isdisc{f})=A\isdisc{f}$; $A\in
\Reals^{N\times N}$. 
Splitting A into the Hermitian, $D$, and skew-Hermitian, $C$, parts \cite{Beilina2017},
%Taking the only possible Hermitian ($D$) and skew-Hermitian ($C$) splitting of $A$ \cite{Beilina2017}
\begin{align}
C&= \displaystyle{\frac{1}{2}\left(A-A^*\right)}; \\
D&= \displaystyle{\frac{1}{2}\left(A+A^*\right)}; \\
A&=C+D,
\end{align}
where $(\cdot)^*$ indicates the conjugate transpose. The previous matrices have interesting 
properties regarding the inner product:
\begin{equation}
 \begin{array}{c}
  \innprod{C \isdisc{\psi}}{\isdisc{\psi}}{} \in \mathbb{I}\\
  \innprod{D \isdisc{\psi}}{\isdisc{\psi}}{} \in \mathbb{R}\\
  \innprod{C
\isdisc{\psi}}{\isdisc{\eta}}{}=-\innprod{\isdisc{\psi}}{C\isdisc{\eta}}{}\\
  \innprod{D
\isdisc{\psi}}{\isdisc{\eta}}{}=\innprod{\isdisc{\psi}}{D\isdisc{\eta}}{},
 \end{array}
\end{equation}
where $\{\isdisc{\psi},\,\isdisc{\eta}\}\in\Comxs^N$.

Both matrices $C$ and $D$ are related, respectively, to the real and the imaginary part of
the $\gamma$ scalars of Eq.(\ref{definitionGamma}). For a given
discretization, these scalars can be numerically calculated: 
\begin{equation}\label{gammaSeparated}
 \begin{array}{c}
  {\isApp{\gamma}_{mn}}=\innprod{A\isdisc{\phi_m}}{\isdisc{\phi_n}}{}\\
  {\isApp{\gamma}_{mn}}^C=Im(\isApp{\gamma}_{mn})=\innprod{C\isdisc{\phi_m}}{\isdisc{\phi_n}}{}\\
  {\isApp{\gamma}_{mn}}^D=Re(\isApp{\gamma}_{mn})=\innprod{D\isdisc{\phi_m}}{\isdisc{\phi_n}}{}.
 \end{array}
\end{equation}

This development holds because $\Phi$ is orthonormal and due to the
nature of the skew and Hermitian operators. They can be calculated from $A$:
\begin{equation}\label{gammaSubs}
 \begin{array}{c}
  {\isApp{\gamma}_{mn}^C}=\innprod{C\isdisc{\phi_m}}{\isdisc{\phi_n}}{}=\displaystyle{\frac{\innprod{A\isdisc{\phi_m}}{\isdisc{\phi_n}}{}-\innprod{\isdisc{\phi_m}}{A\isdisc{\phi_n}}{}}{2}}\\
  {\isApp{\gamma}_{mn}^D}=\innprod{D\isdisc{\phi_m}}{\isdisc{\phi_n}}{}=\displaystyle{\frac{\innprod{A\isdisc{\phi_m}}{\isdisc{\phi_n}}{}+\innprod{\isdisc{\phi_m}}{A\isdisc{\phi_n}}{}}{2}},
 \end{array}
\end{equation}
where the values of $\gamma_{mn}$ are the different elements of the matrix $\Gamma$.
To simplify the analysis, we propose to compute the root mean 
square of the second index of the values of $\gamma$,
\begin{equation}\label{rmsLambda}
 \begin{array}{c}
  {\isApp{\lambda}_{m}^C}=\sqrt{\sum_{n}
\displaystyle{\left({\frac{\innprod{A\isdisc{\phi_m}}{\isdisc{\phi_n}}{}
-\innprod{\isdisc{\phi_n}}{A\isdisc{\phi_m}}{}}{2}}\right)^2}}\\
  {\isApp{\lambda}_{m}^D}=\sqrt{\sum_{n}
\displaystyle{\left({\frac{\innprod{A\isdisc{\phi_m}}{\isdisc{\phi_n}}{}
+\innprod{\isdisc{\phi_n}}{A\isdisc{\phi_m}}{}}{2}}\right)^2}}.
 \end{array}
\end{equation}

This procedure allows a faster comparison between the analytical value and the recovered numerical ones. Thus, if the 
recovered ${\isApp{\lambda}_{m}^D}$ is not null, the differential operator will have diffusive
behavior. If the ratio between $|{\isApp{\gamma}_{mm}}|$ and
${\isApp{\lambda}_{m}^C}$ is lesser than the unity,
then values of ${\isApp{\gamma}_{mn}}$ which should be zero will have a non-zero value. And finally, the ratio between 
${\isApp{\gamma}_{mm}}$ and the reference parameter indicate the deviation of the numerical discretization 
from its expected value.

In summary, given a set of approximations of differenciating operators
$\{A_1,\,A_2,\,\ldots,\,A_P\}$,
their dispersion properties on a representative mesh can be compared
after computing the quantites $\isApp{\lambda}_{m}^{C_j}$ and
$\isApp{\lambda}_{m}^{D_j}$ for each element $\phi_m$ of an
orthonormal basis and discrete operator $A_j$. The following subsections address
 the evaluation of $A_j$ and the selection of an appropiate orthonormal basis
$\Phi$. 

%% file: 4.GradientEval.tex
\label{sec:Method:subsec:Dimms}
To evaluate the approximations of the derivative operators we propose to
substitute Eq.(\ref{eq:WaveEq}) with
\begin{equation} \label{eq:ConvEq}
\frac{\partial r_j}{\partial t} + C(c_j,\,r_j)=0,\;\qquad j\in[1,\,D+2]; 
\end{equation}
where $C(c_j,\,r_j)$ is the convection operator on $r_j$. After this, approximations of
derivative operators can be easily obtained with $c_j=1$. Notice that in the
continuous space or with constant $c_j$ Eqs. (\ref{eq:WaveEq}) and
(\ref{eq:ConvEq}) are equivalent. 
%In addition, discrete convection operators are always discretized on the same basis $\Omega$ of
%the magnitudes they are applied to to allow for time integration. 
In practical simulations, the discrete convective operator is an isomorphism,
i.e. $C:\Omega \mapsto \Omega$.
Hence, this resolves the eventual problem of the different basis pointed 
out in former paragraphs of this section.
Furthermore, Eq.(\ref{eq:ConvEq}) is
advantageous in discretizations where $c_j$ is not constant. In fact, in
the Euler Equations in form of Eq.(\ref{eq:ConvEq}), $c_j:\Reals^{D+2}\mapsto\Reals$; $c_j(\isdisc{r})$
and their spatial variations can only be neglected for small
perturbations (e.g. acoustics). 

However, this assumption does not hold when simulating hydrodynamic regions. In these regions,
the spatial variations of $\isdisc{r}$ are of the same order of
magnitude and spatial variatons of $c_j$ cannot be neglected.
Nonetheless, the literature shows how low-dispersion arguments based on
the classical methodology have been employed to support using high-order
numerical schemes in these regions. We expect that the dispersion
analysis developed here with the derivatives computed via Eq.(\ref{eq:ConvEq}) 
will allow comparing schemes in hydrodynamic regions.

%Until here, all the developments have been made on 1D and approximations
%to the derivative. The
%considerations in section \ref{sec:AnalDerivation} can be extended to
%multiple dimmensions straighforwardly. Regarding the methodology, one
%faces two main challanges arising from derivatives or gradient operators not being
%homeomorphisms. The first occurs when the basis $\Omega$ on which a function $f$ is
%discretized does not coincide with the basis of functions $\Omega'$ on
%which its derivative is discretized. For example, in FV formulations on
%structured grids,
%one typically defines gradients at cells interfaces, and there is one
%$\Omega'$ per dimmension while on unstructured mehses, this approach
%does not apply. The second arises due to the simple fact that, even when
%$\Omega=\Omega'$, To ease the 
%extended to various dimmensions by \textcolor{red}{Explicar coses de
%dimensions i tal} and \textcolor{red}{Com evaluar gradients a partir de
%Convectius i per que el problema de $\Omega\neq\Omega'$ esta resolt}.

%% file: 5.OrthonormalBasis.tex
\label{sec:Method:subsec:Orthonormal}
The method described above has to be applied on an appropriate orthonormal
basis of the discrete fields, $\Phi$.
The classical analysis \cite{Lele1992,Tam2012, Tam1993}
performs the dispersion error analysis on uniform structured meshes. On
them, it is straightforward to use the projections of sinusoids on
the canonical basis $\Omega^0$, i.e., define a change of basis of the type
$B:\Reals^N\mapsto\Reals^N$. For example, in Finite Differences or
Finite Volumes formulations, one can use
 $b_{jk}=\innprod{\sin(jx)}{\omega^0_k}{\Omega_x}$. However, in non-uniform or unstructured meshes,
this procedure does not generate orthonormal basis.

For these cases, we propose to use the eigenvectors of discrete Laplacian operators defined
on such meshes. This basis is used in signal
analysis and related fields (see Shuman et al.\cite{Shuman2013}).
Among others, the properties of the eigenvectors of the discrete Laplacian operator are:
\begin{itemize}
 \item {The eigenvectors are orthonormal.} 
 \item {In evenly spaced domains, the eigenvectors are discretized sinusoids.}
 %\item {In the continuous case, i.e. infinite oscillation modes, the eigenvectors are waves and the eigenvectors wavelengths.}
 \item {In the continuous limit, its eigenvectors and eigenvalues collapse into 
 the eigenfunctions of its continuous counterpart, i.e. sinusoids.}
 \item {They retain the concept of mesh connectivity.}
\end{itemize}

Furthermore, 
\begin{equation}
 L=G^*G,
\end{equation}
where $L\in\Reals^{N\times N}$ is a discrete Laplacian,
$G\in\Reals^{M\times N}$ a discrete gradient (see \cite{Bochev2006}) and
$G^*$ the conjugate-transpose of $G$. It is important to notice that $G$ is not necessarily the
differenciating operator on which the dispersion error analysis is to be
conducted but a differencing operator that holds the equality. 
Actually, $\Phi$ should be independent of the scheme under
study to allow comparisons if several of them will be tested.

The Singular Value Decomposition of $G$ relates its right-eigenvectors
$\{\isdisc{g_1},\,\isdisc{g_2},\,\ldots,\,\isdisc{g_N}\}$
and singular values $\{g_1,\,g_2,\,\ldots,\,g_N\}$ with the eigenvectors
$\{\isdisc{l_1},\,\isdisc{l_2},\,\ldots,\,\isdisc{l_N}\}$ and
eigenvalues of $\{l_1,\,l_2,\,\ldots,\,l_N\}$
of $L$. Specifically, $g_n=\sqrt{l_n}$ and $\isdisc{g_n}=\isdisc{l_n}$.

Thus, any scheme under analysis will be compared to a reference
gradient. With this, we propose to calculate the quantities of Eq.(\ref{rmsLambda})
 with $\Phi=
\{\isdisc{l_1},\,\isdisc{l_2},\,\ldots,\,\isdisc{l_N}\}$ and
$\lambda_j=\sqrt{l_j}$ for all $j\in[1,\,N]$.

\begin{figure}
 \centering
 \includegraphics[width=0.4\textwidth]{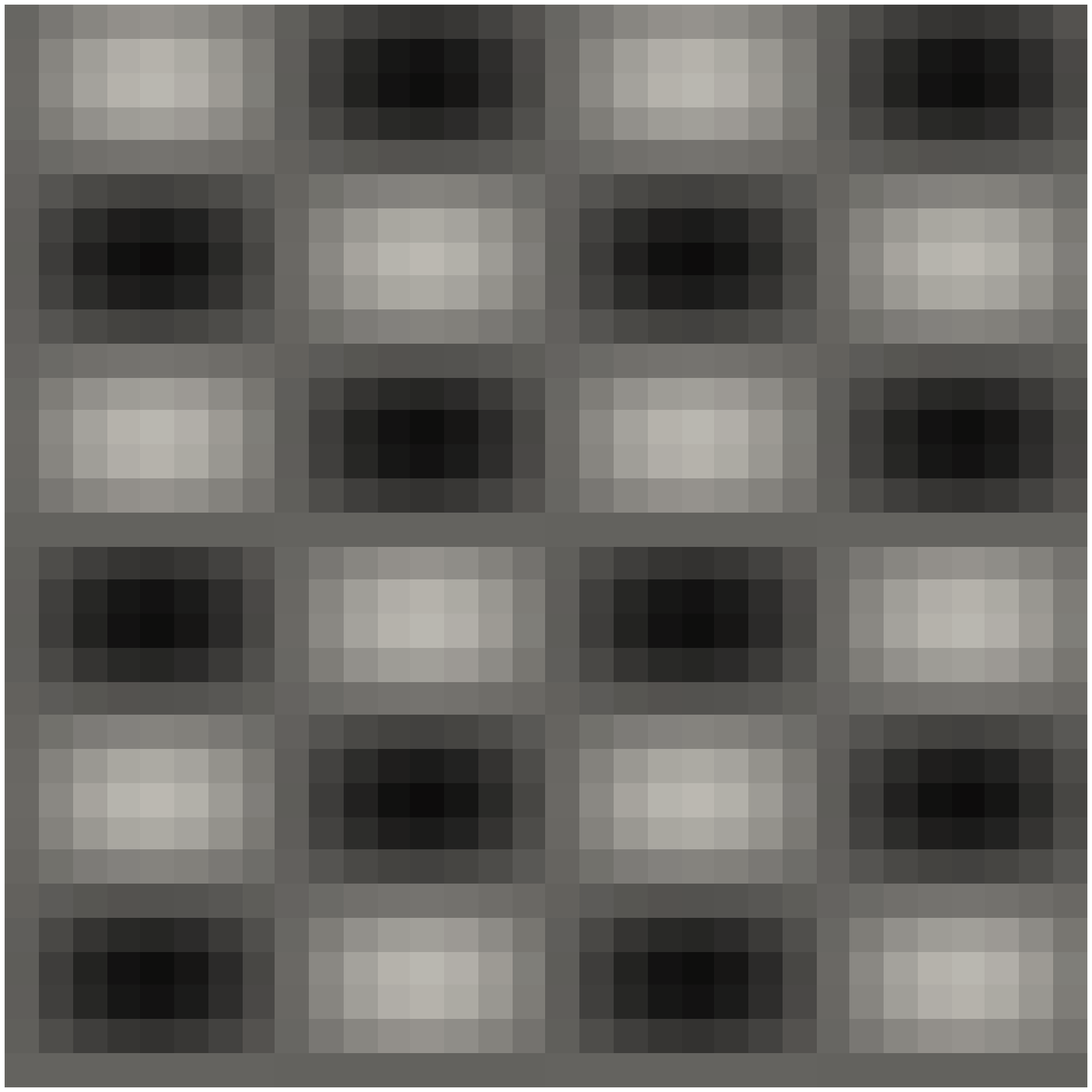} 
 \includegraphics[width=0.4\textwidth]{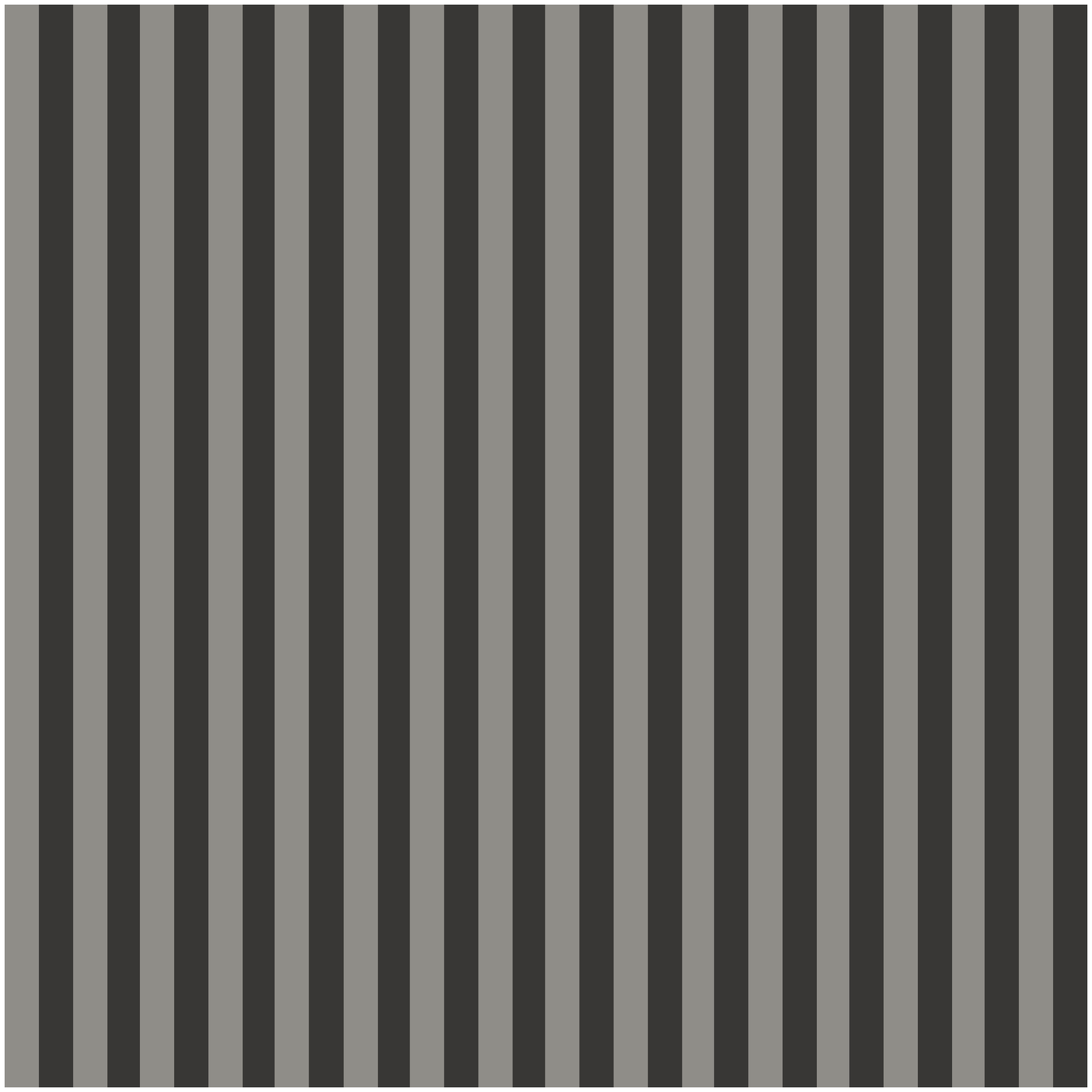} 
 \captionof{figure}{Examples of discrete eigenvectors in two-dimensional structured uniform mesh.
 Left: low eigenvalue associated, right: high  eigenvalue associated.} \label{eigenStUn} 
\end{figure}
\begin{figure}
 \centering
 \includegraphics[width=0.4\textwidth]{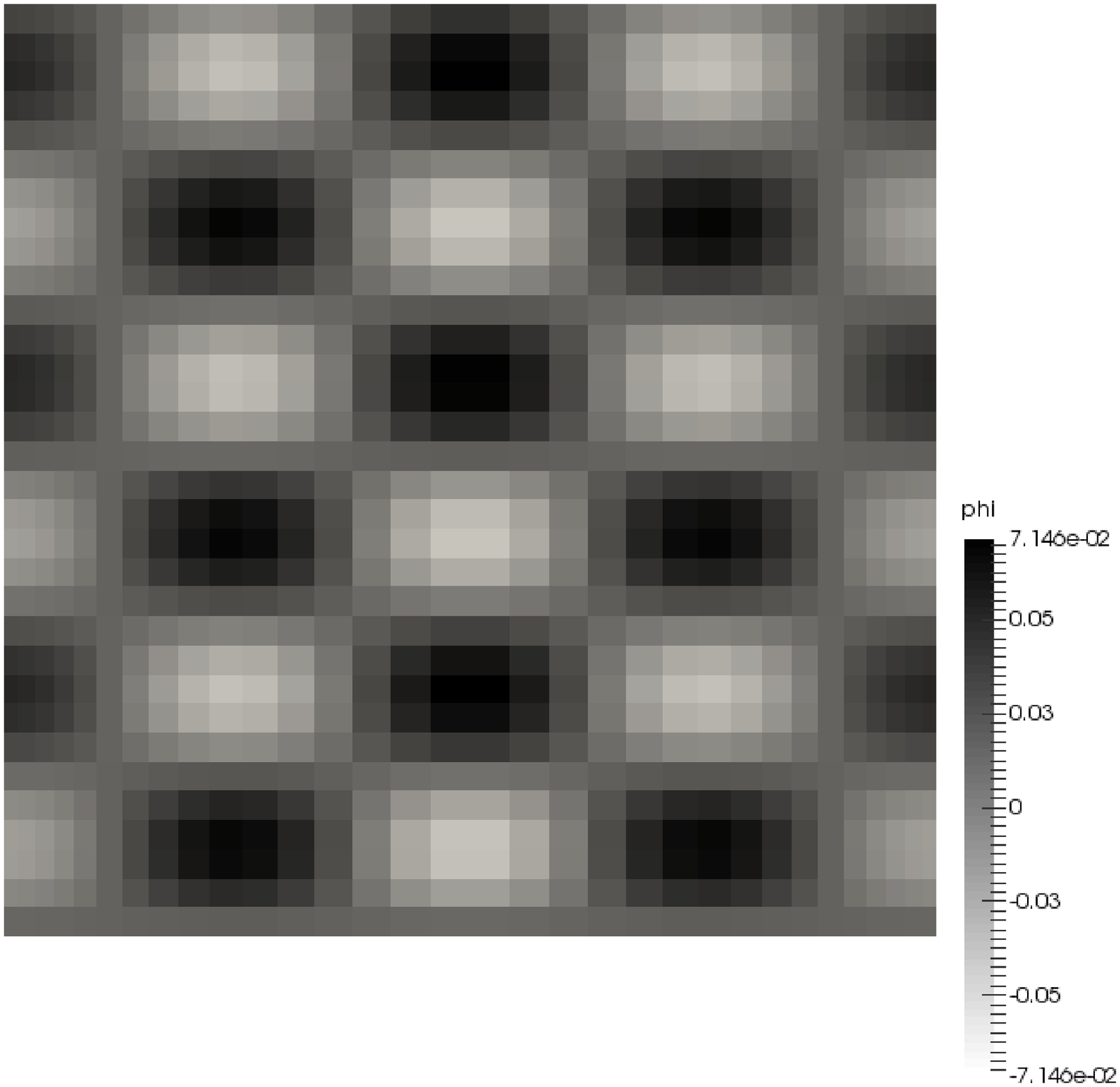} 
 \includegraphics[width=0.4\textwidth]{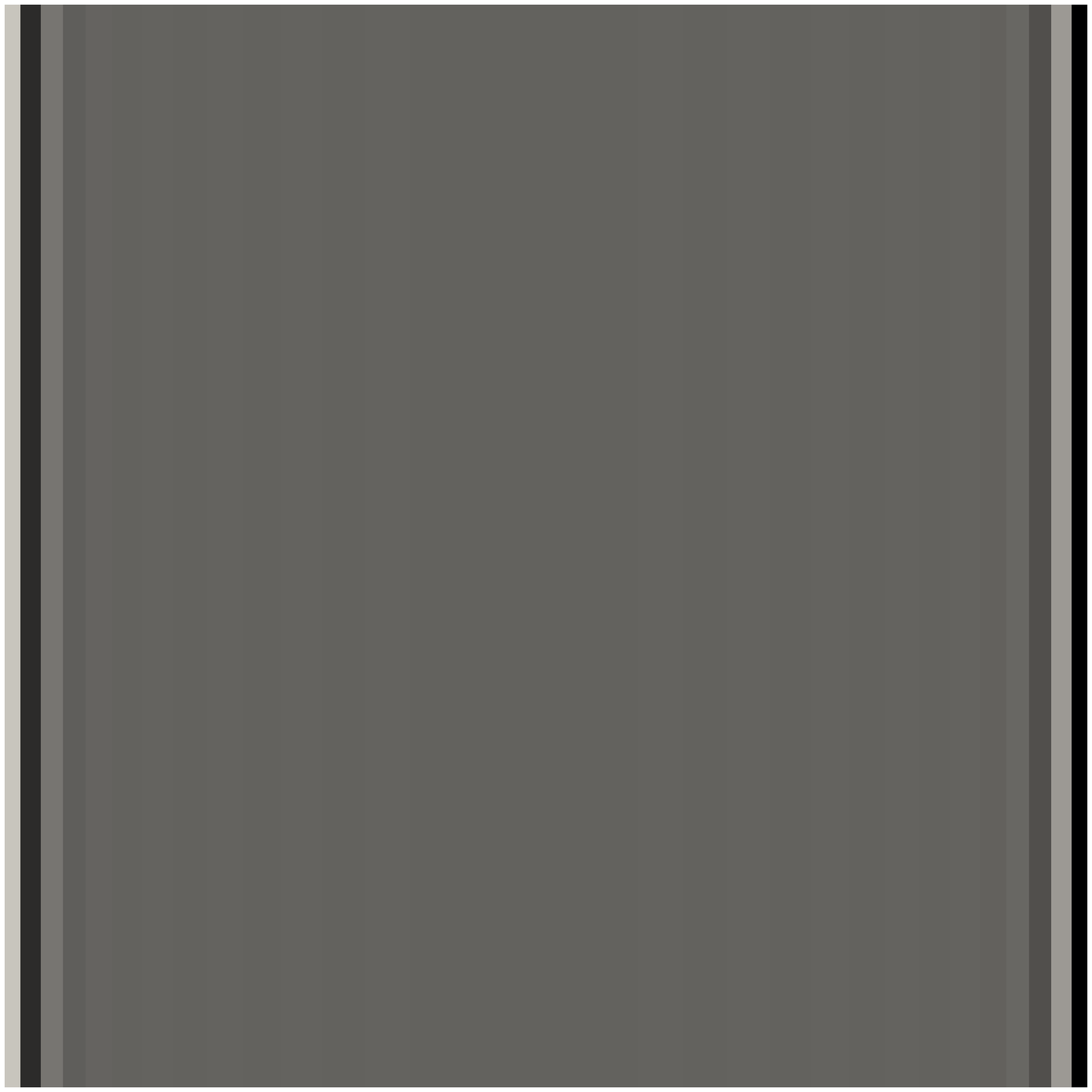} 
 \captionof{figure}{Examples of discrete eigenvectors in two-dimensional structured non-uniform mesh.
 Left: low eigenvalue associated, right: high eigenvalue associated.} \label{eigenStNUn} 
\end{figure}
\begin{figure}
 \centering
 \includegraphics[width=0.4\textwidth]{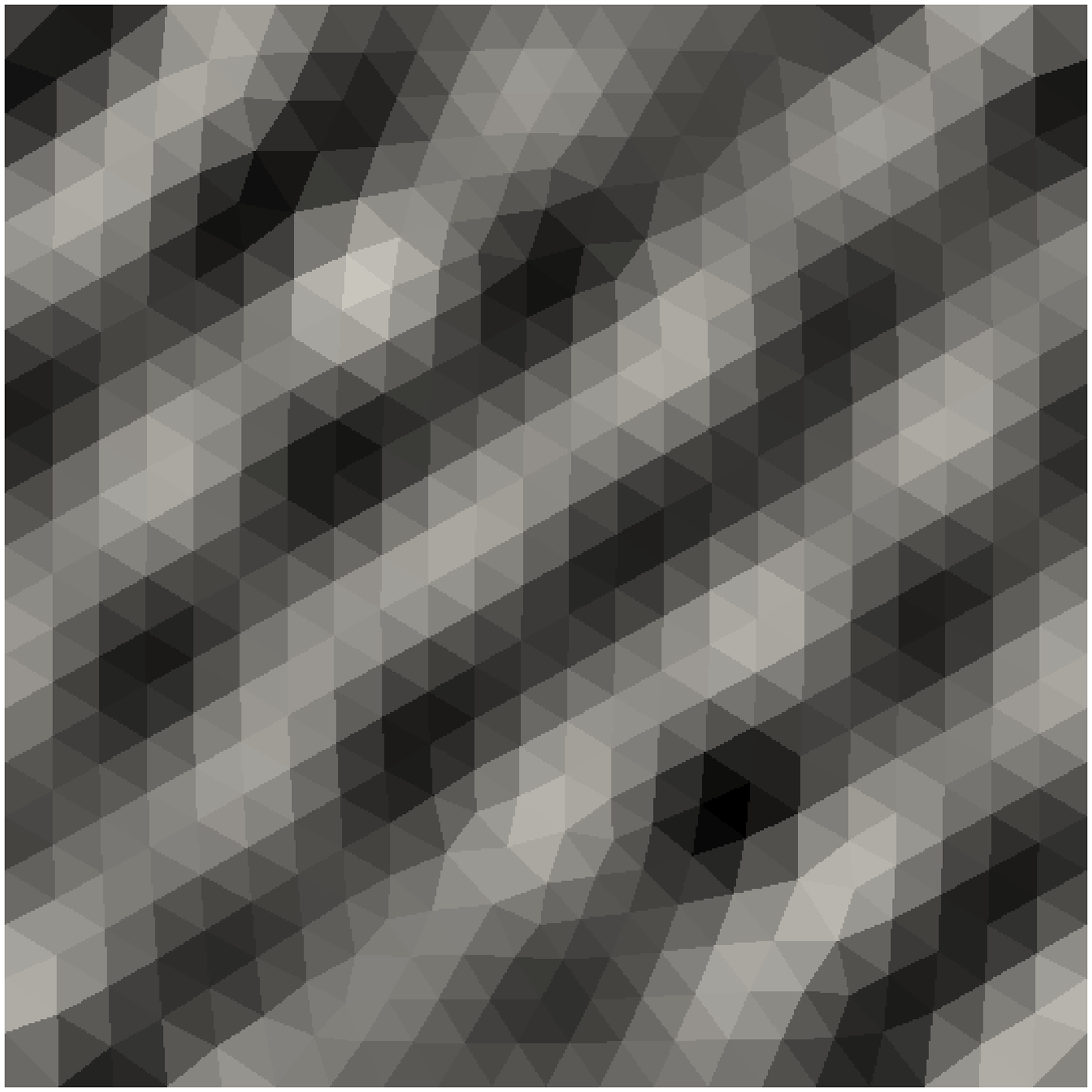} 
 \includegraphics[width=0.4\textwidth]{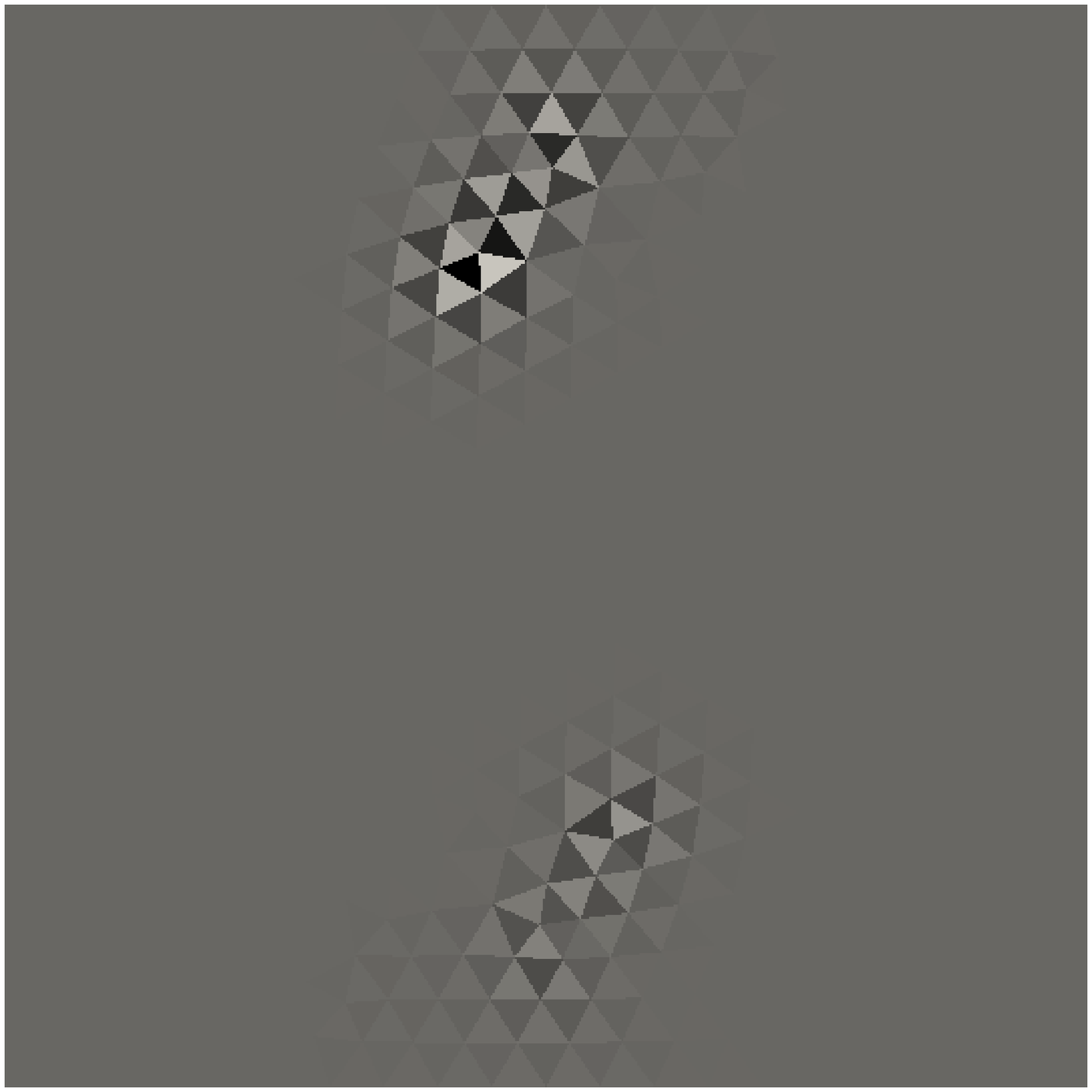} 
  \captionof{figure}{Examples of discrete eigenvectors in two-dimensional unstructured mesh.
  Left: low eigenvalue associated, right: high eigenvalue associated.} \label{eigenUn} 
\end{figure}
The discrete eigenvectors of a 2nd order discrete symmetric Laplacian operator are shown in 
figures \ref{eigenStUn}-\ref{eigenUn}. To ensure symmetry of the obtained matrix, 
this Laplacian has been constructed as:
\begin{equation}\label{discLap}
 L=\sum_{p\in Nb}a_p(\phi_p-\phi_o)=0;\;\qquad a_p=\frac{A_{op}}{\vv{d_{op}}\vv{n_{op}}\displaystyle {\frac{V_p+V_o}{2}}},
\end{equation}
where $A_{op}$ is the intersection area between volumes $o$ and $p$, $\vv{n_{op}}$ its normal vector, 
$\vv{d_{op}}$ the distance between centroids and $V_p$ and $V_o$, the volumes $p$ and $o$. 
$Nb$ indicates the number of neighbours that surround control volume $o$.

The results resemble, into some extent, a discrete sinusoid; this is more obvious when low modes are selected. 
When higher modes are selected, as on the figures on the right, 
this similarity can be lost if non-uniform, or non-structured meshes are used.

%% file: 6.ChangeBasis.tex
It may be argued that the present methodology is that the discrete
eigenvalues of $L$ are not directly related with waves in the physical,
continuous, space. However, the closer the discrete $L$ approximates the
continuous $\nabla^2$, the closer are its eigenvalues and eigenvectors
to resemble physical waves. Put short, using high
order $L$ resolves the problem, at least for long wavelengths with
respect to the mesh characteristic spacing. But still,
the user of the present methodology does not have a strict control
of the wavelengths that will be the eigenvalues of $L$ on a mesh.

The following method allow to evaluate dispersion errors for
any function $f$ and, in particular, a sinusoid with wavenumber 
``$k$'', e.g., $f(k,\,x)=\sin(kx)$.
Expressing $f$ on $\Omega$, $f(k,\,x)\simeq\sum_m a(k)_m\phi_m(x)$, with
$a(k)_m=\innprod{f}{\phi_m}{\Omega_x}$, which are the 
coordinates of the sinusiod function in the eigenvectors basis. The
derivative is approximated with
\begin{equation}
 f'(k,\,x)\simeq\frac{d}{dx}\sum_m a(k)_m \phi_m(x)=\sum_m a(k)_m
\phi_m'(x)= \sum_m a(k)_m
\sum_n\gamma_{mn} \phi_n(x),
\end{equation}
%where, $\gamma_{mn}$ is as in Eq.({\ref{definitionGamma}). 
%\begin{equation}
% f=A\cdot\Phi
%\end{equation}
%And:
%\begin{equation}
% f'=A\cdot\Gamma\cdot\Phi
%\end{equation}
%Where the terms $f$ and $f'$ are vectors of $n$ elements, containing each element the distribution in $x$.
%Matrix $A$ are the terms $a(n,k)$ ordered in a matrix form. $\Phi$ is the vector of $k$ elements containing
%each element the distribution in $x$ of the original basis. Finally, $\Gamma$ are the terms $\gamma(m,n)$ 
%presented in equation \ref{definitionGamma}.
We can compute an approximation to the classical dispersion error: 
%Following the same methodology as in equation \ref{definitionGamma}, where the derivative is projected onto
%the primitive function:

\begin{equation}\label{dispersionSinusoid2}
 S(k,l)=\innprod{\isApp{f'}(k,\,x)}{f(l,\,x)}{\Omega_x}
%\innprod{f'(k,\,x)}{f(k,\,x)}{\Omega_x}
 \simeq \innprod{\isApp{\Gamma}\isdisc{a}(k)}{\isdisc{a}(l)}{}.
%\innprod{f'(k,\,x)}{f(k,\,x)}{\Omega_x}  
% A\cdot\Gamma\cdot\Phi,A\cdot\Phi>=A\cdot\Gamma\cdot\Phi\cdot(A\cdot\Phi)^T=
 %A\cdot\Gamma\cdot\Phi\cdot\Phi^T\cdot A^T=A\cdot\Gamma\cdot I\cdot A^T
\end{equation}
where $\isApp{\Gamma}$ is defined as in Eq.(\ref{gammaSubs}).
As a particular case, when $l$ is equal to $k$, the above definition becomes
the usual expression of dispersion error on evenly distributed meshes, i.e.
how the derivative of a mode projects in the same mode:
\begin{equation}\label{dispersionSinusoid}
 S(k,k)=\innprod{\isApp{f'}(k,\,x)}{f(k,\,x)}{\Omega_x}
%{ \innprod{f'(k,\,x)}{f(k,\,x)}{\Omega_x}}
 \simeq \innprod{\isApp{\Gamma}\isdisc{a}(k)}{\isdisc{a}(k)}{}.
%\innprod{f'(k,\,x)}{f(k,\,x)}{\Omega_x} }, 
% A\cdot\Gamma\cdot\Phi,A\cdot\Phi>=A\cdot\Gamma\cdot\Phi\cdot(A\cdot\Phi)^T=
 %A\cdot\Gamma\cdot\Phi\cdot\Phi^T\cdot A^T=A\cdot\Gamma\cdot I\cdot A^T
\end{equation}

%Where $\Phi\cdot\Phi^T$ is equal to the identity matrix due to the fact $\Phi$ is an orthonormal basis.\\
%It is important to notice that equation \ref{dispersionSinusoid} is not a basis change; just if matrix $A$
%is orthogonal this equality will represent a basis change. When uniform meshes are used, discrete sinusoids 
%form an orthogonal basis and consequently the matrix A becomes orthogonal too. However, in a more general
%case with non-uniform or non-structured meshes, a set of discrete sinusoids do not form an orthogonal basis.\\
Thus, the presented methodology is able to numerically approximate the
results of classical analysis on general meshes, whereas the classical
analysis cannot. The same numerical procedure allows to study 
the projection of the selected eigenvectors into any function of interest.

%% file: 7.ApplicationCase.tex
The set of numerical experiments reported here illustrates the use of
the developed methodology.
We focus on stretched structured meshes as this type of meshes is a
common practice that allows limiting the total mesh sizes and the
associated computing cost and time in CAA. It allows, among others, a controlled refinement near walls 
or zones where a smaller mesh size is required without compromising the total amount 
of mesh nodes. However, a variation of the mesh size introduces errors
that can potentially lead to instabilities.
%; more precisely, the mesh growth rate is a known source of instabilities or numerical errors.
%fiquem referencies Shur%
Estimating the magnitude of these errors before running a simulations
allows designing of the simulations with the desired precision and
accuracy.

The selected experiments consist of comparing dispersion errors of
approximate derivatives on non-uniform structured 1D meshes with increasing
degree of constant volume growth ratio. The test functions used
to compute dispersion error are the discrete eigenvectors computed from the discrete Laplacian. 
The Laplacian is discretized as in Eq.(\ref{discLap}), following a second order approximation of the differential operator, ensuring
symmetry of the obtained matrix, and with periodic boundary conditions. 
Once constructed, the eigenvalues and eigenvectors of the matrix, 
the discrete Laplacian matrix, are calculated by means of QR reduction algorithm present in 
GNU scientific library package \cite{Galassi2007}. The obtained test functions are then convected by using a 
second-order Symmetry Preserving approach of the discrete transport term of Navier Stokes equations. 
This term is linearized, i.e. constant advection velocity, in order to compare
the presented method with the classical one due to the fact the latter implies a linear 
convective term. This means the studied convective term in one direction is:
\begin{equation}
C((1,0,0),\, \isdisc{\phi})_{i}= \frac{d}{dx} \isdisc{\phi}_i.
\end{equation}
And its discrete counterpart in a Finite Volume framework:
\begin{equation}
\frac{d}{dx} \isdisc{\phi}_i\simeq{\frac{1}{V_i}\sum_{f\in V_i}\phi_f A_f n_x}.
\end{equation}

Due to the fact a second-order Symmetry Preserving discretization has been selected, the value
$\phi_f$ is equal to $\frac{1}{2}(\isdisc{\phi}_i + \isdisc{\phi}_{nb})$.
Once advected, the matrix $\hat{\Gamma}$ is constructed following Eq.(\ref{gammaSubs}). In 
this case, having selected a conservative and non-diffusive discretization, the terms
related to ${\isApp{\gamma}_{mn}}^D$ are null. The final step of the procedure consists in
calculating the dispersion characteristics in the sinusoids base by means of Eqs.(\ref{dispersionSinusoid}) and
(\ref{dispersionSinusoid2}). These values constitute the elements of
matrix $S\in\Comxs^{N\times N}$.

\begin{figure}
 \centering
 \includegraphics[width=0.6\textwidth]{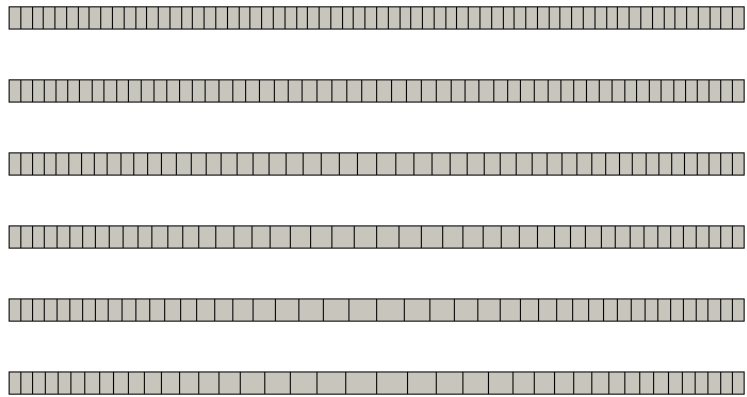} 
 \captionof{figure}{Some of the used meshes. From top to bottom: from uniform mesh to 
 5\% stretching with increments of 1\%. } \label{meshesStretch} 
\end{figure}

The meshes for the tests are 1D stretched meshes. 
The total size of the domain is the same in all the meshes and tests.  
The errors commited with the stretched meshes are compared with those commited with
a uniform, cartesian 1D mesh. The range of studied wavenumbers is
$\nu=[0:k_{Max}]=[0:2\pi/\lambda_{Min}]$; being $\lambda_{Min}$ the minimum
studied wavelength and equal to $2\Delta x_{Min}$. $\Delta x_{Min}$ the minimum cell size of the mesh.
The volume size of the uniform mesh is the minimum volume size of the stretched ones; thus all meshes
share the same maximum studied wavenumber. The characteristics of these meshes are reported in table 
\ref{table:meshCharac} and some of them are shown in figure \ref{meshesStretch}.

\begin{table}
  \centering
\begin{tabular}{ccccc} \\
\toprule
$\Delta x_{Min}$ & $\Delta x_{Max}$ & Volume growth ratio & Number of volumes & Mesh savings [\%] \\  
\midrule
0.015625 & 0.015625 & 1.0 & 64 & 0\\
0.015625 & 0.018147 & 1.005 & 60 & 6.25\\
0.015625 & 0.020645 & 1.01 & 56 & 12.5\\
0.015625 & 0.023356 & 1.015 & 54 & 15.625\\
0.015625 & 0.025634 & 1.02 & 50 & 21.875\\
0.015625 & 0.028261 & 1.025 & 48 & 25\\
0.015625 & 0.030837 & 1.03 & 46 & 28.125\\
0.015625 & 0.033305 & 1.035 & 44 & 31.25\\
0.015625 & 0.037030 & 1.04 & 44 & 31.25\\
0.015625 & 0.039379 & 1.045 & 42 & 34.375\\
0.015625 & 0.041458 & 1.05 & 40 & 37.5\\ 
\bottomrule
\end{tabular}
\caption[Mesh characteristics]{Mesh characteristics.}
\label{table:meshCharac}
\end{table}

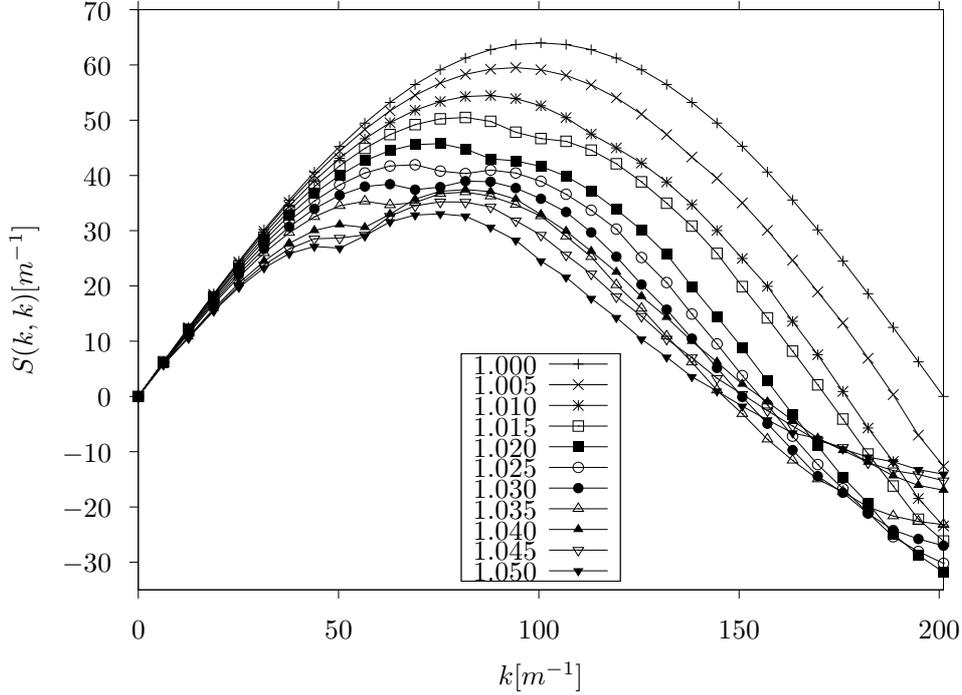
\begin{figure}
 \centering
 \resizebox{0.8\textwidth}{!}{\input{eigen2SinDiag.tex}}
 \caption{Diagonal terms vs imposed wavenumber in sinusoids space for different ratios.}\label{eigen2SinDiag}
\end{figure}

Figure \ref{eigen2SinDiag} approximates the classical analysis: it shows the diagonal part of the recovered 
$S$ matrix. Due to the fact a Symmetry Preserving discretization has been used these values are purely imaginary.  
For the uniform mesh, the analysis gives the same results as the classical analysis of Tam \cite{Tam1993}
and Lele \cite{Lele1992}. This was expected because for structured meshes with periodic boundary conditions the
discrete gradient operator can be expressed as a circulant matrix. Eigenvectors of these matrices are known
to be discrete sinusoids. The results of the stretches meshes provide new insight in dispersion errors as 
this kind of result is innovative. 

First, for the selected range of wavenumbers the stretched meshes exhibit reflections, or a change
in the sign of the propagating wavenumber, at mid-to-higher wavenumbers. These reflections occur
before, in a lesser wavenumber, in highly stretched than in slightly stretched meshes.
Another immediate result extracted from figure \ref{eigen2SinDiag} is that stretched meshes 
introduce an intermediate frequency on the positive side of the plot at which an inflection 
point appears. Later it will be discussed the relation of this kind of frequency with 
dimensional aspects of the mesh.

%Additionally, if we plot the numerical wavenumber for a structured uniform mesh with 
%the same number of control volumes as the stretched mesh another feature is shown.
%the same results are plotted, just for uniform and 5\% stretched meshes, in Figure
%\ref{eigen2SinDiagDetail}. Increasing the volume growth ratio distorts the shape of the plot in 
%comparison with a uniform mesh. This happens because even with the same 
%number of cells, the maximum volume size of a stretched mesh is larger
%than that of the uniform mesh. On the other
%hand, plotting the numerical wavenumber of a uniform mesh with the same
%volume size as the the maximum of one of the stretched meshes, the
%analysis shows a poorer performance for this coarse uniform mesh.

%\begin{figure}
% \centering
% \resizebox{0.8\textwidth}{!}{\input{images/eigen2SinDiagDetail.tex}}
% \caption{Diagonal terms vs imposed wavenumber in sinusoids space for maximum and minimum ratio.
% Comparison with structured meshes with same number of control volumes and spacing equal to the 
% maximum one in the stretched mesh}\label{eigen2SinDiagDetail} 
%\end{figure}

\begin{figure}
 \centering
 \resizebox{0.8\textwidth}{!}{\input{eigen2SinLam.tex}}
 \caption{$\widetilde{\lambda}$ vs imposed wavenumber in sinusoids space for different ratios.}\label{eigen2SinLam}
\end{figure}
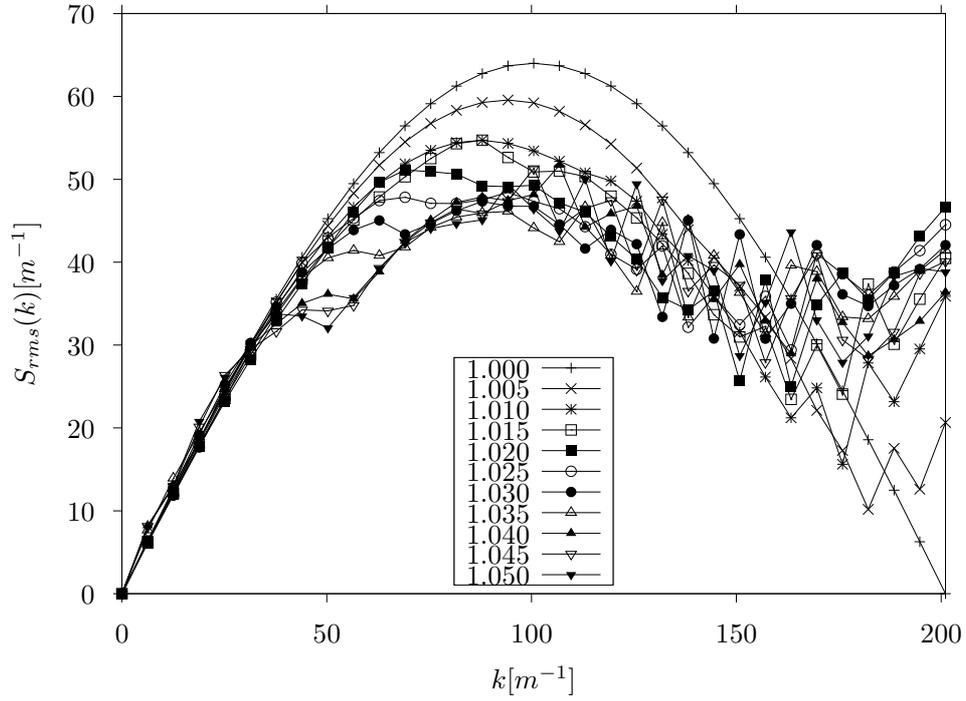

Figure \ref{eigen2SinLam} includes the effects of the off-diagonal terms of $S$, where 
the root mean square of the row elements is presented. This parameter is the 
equivalent of $\lambda(m)$ presented in Eq.(\ref{rmsLambda}) but in the space of sinusoids.
As it can be seen, the shape of the plots is different from those in figure \ref{eigen2SinDiag};
these differences are enlarged at higher strecthing ratios. If figures \ref{eigen2SinLam} and 
\ref{eigen2SinDiag} are different this means that other elements than the diagonal appear. 
Therefore, the recovered modes are composed of various frequencies instead of a single one, 
which should be only the diagonal term.

How the recovered mode, using the presented method, is composed of different modes is illustrated 
in figures \ref{aliasingMatrix1000} and \ref{aliasingMatrix1050}, the uniform and mesh stretched a 5\%. 
In the figures we use modes instead of frequencies, distinguishing between sinus and cosinus in order to show that 
even modes, i.e. sinus, project onto odd modes, cosinus, and viceversa. Ideally, the matrices of the figures 
should be diagonal, mode '$j$' derivated projects into mode '$j+1$' or '$j-1$' and viceversa, and skew-symmetric.

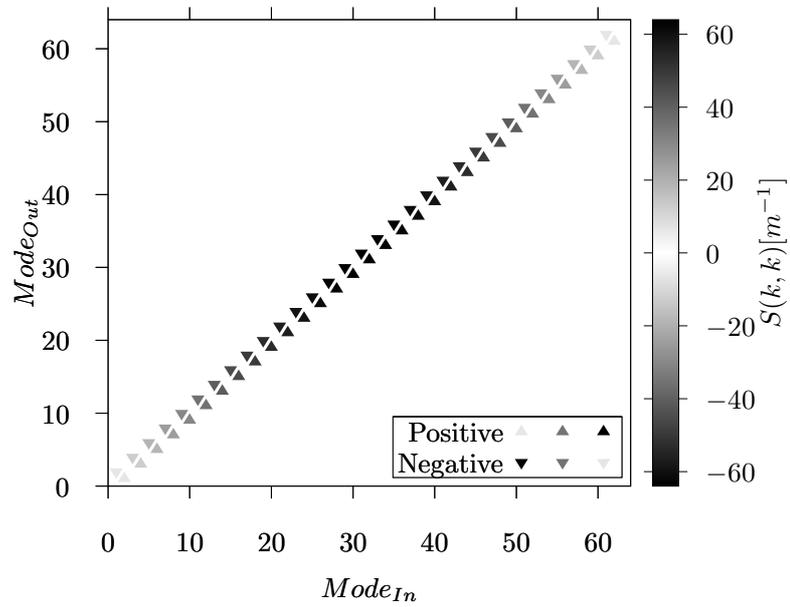
\begin{figure}
 \centering
 \resizebox{0.8\textwidth}{!}{\input{aliasingMatrix1000.tex}}
 \caption{Dispersion matrix in sinusoids space for uniform mesh. Input mode in the horizontal axis
 and recovered mode of the derivative in the vertical axis.}\label{aliasingMatrix1000}
\end{figure}
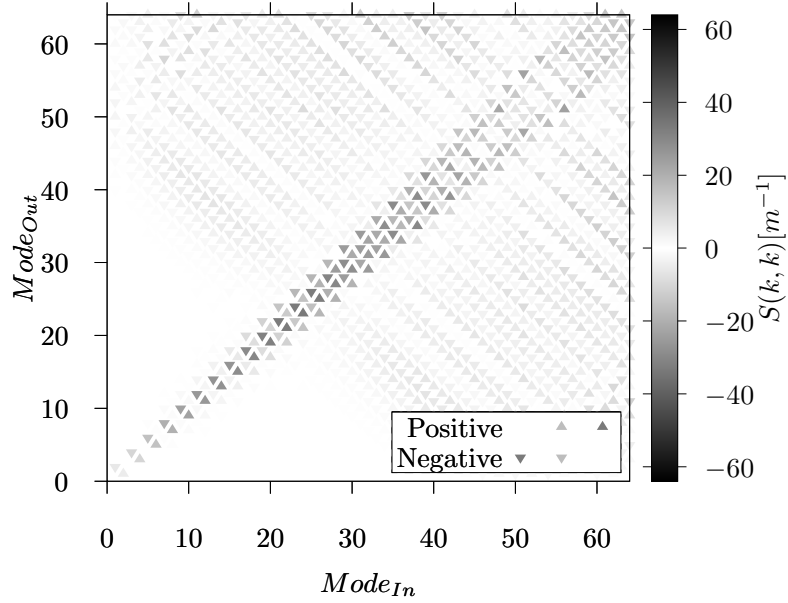
\begin{figure}
 \centering
 \resizebox{0.8\textwidth}{!}{\input{aliasingMatrix1050.tex}}
 \caption{Dispersion matrix in sinusoids space for stretched mesh (ratio 1.05). Input mode in the horizontal axis
 and recovered mode of the derivative in the vertical axis.}\label{aliasingMatrix1050}
\end{figure}

As can be seen, only the uniform mesh provides a diagonal matrix. Stretched meshes provide additional off-diagonal
terms. This means that the modes obtained when deriving in stretched meshes are composed of various modes instead
a single one. The presence of these off-diagonal terms also explains why figures \ref{eigen2SinDiag} and 
\ref{eigen2SinLam} are not identical.

In order to help understanding the mechanisms that happen during all the process, 
we include two new different plots:
\begin{itemize}
 \item First, the equivalent of figures \ref{eigen2SinDiag} and \ref{eigen2SinLam} but in the space of eigenvectors
 instead of the space of sinusoids, i.e. the values of Eq.(\ref{rmsLambda}). This will allow to study just the 
 derivation process without involving a change of basis.
 \item Some values of $\isdisc{a}(k)$ at different meshes, which are the coordinates the sinusiod function 
 in the eigenvectors basis, i.e. how a sinus of specific frequency is composed of several eigenmodes. This will
 help to understand how frequencies are composed of the different eigenmodes.
\end{itemize}

Previously it has been commented the existence of an intermediate frequency in figure \ref{eigen2SinDiag}.
It is difficult to relate the frecuency in this plot, and even quantify it, to dimensional aspects of the mesh.
However, in figure \ref{eigen2eigenDiag} this frequency, or eigenvalue in this figure, is more easily quantified. 
As can be seen, all the curves present a sudden decay and then a stabilization. The eigenvalue at which the 
decay transitions into stabilization has been found to be equal to $\frac{2}{\Delta x_{Max}}$ for each mesh.

\begin{figure}
 \centering
 \resizebox{0.8\textwidth}{!}{\input{eigen2eigenDiag.tex}}
 \caption{Diagonal terms vs imposed eigenvalue in eigenvectors space for different ratios.}\label{eigen2eigenDiag}
\end{figure}
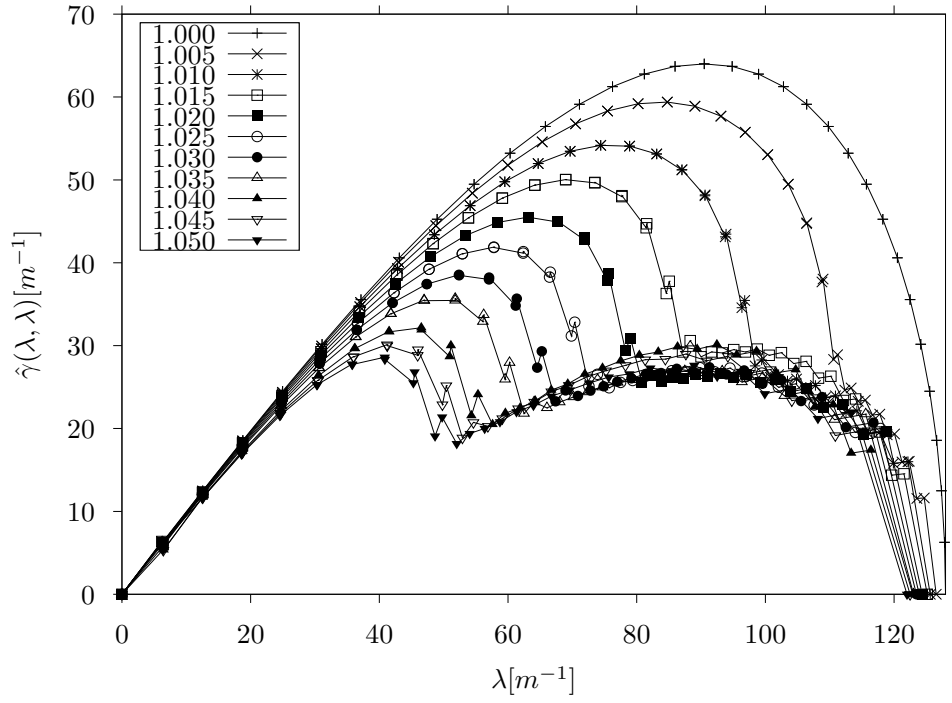
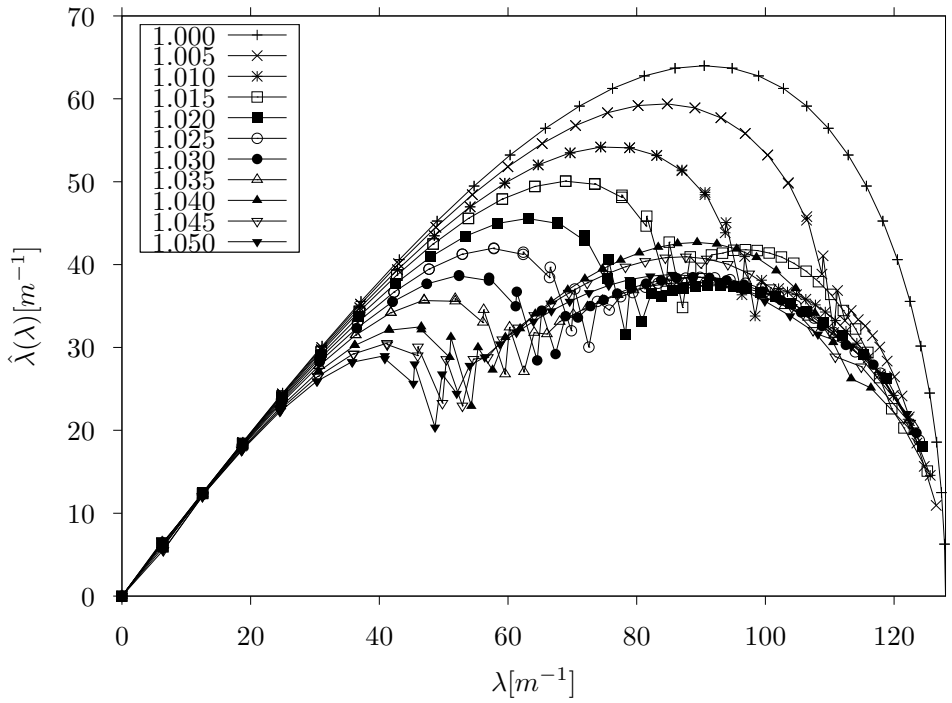
\begin{figure}
 \centering
 \resizebox{0.8\textwidth}{!}{\input{eigen2eigenLam.tex}}
 \caption{$\widetilde{\lambda}$ vs imposed eigenvalue in eigenvectors space for different ratios.}\label{eigen2eigenLam}
\end{figure}

In figure \ref{eigen2eigenLam} the value of $\lambda$ of Eq.(\ref{rmsLambda}) is plotted. This figure 
is slightly different from figure \ref{eigen2eigenDiag} for stretched meshes. This means that 
off-diagonal terms appear in matrix $\Gamma$; thus eigenmodes recovered after derivating are composed 
of various input eigenmodes instead of a single one. Consequently, one of the reasons because figure 
\ref{aliasingMatrix1050} is not diagonal occurs during the derivation process. 

The other reason which will produce non-diagonal matrices will be that during the projection onto the sinusoid
space the operation introduces off-diagonal terms. A single frequency composed of very different eigenmodes will
propagate the information of these modes in a very sparse way, i.e. off-diagonal terms when computing $S(k,l)$.

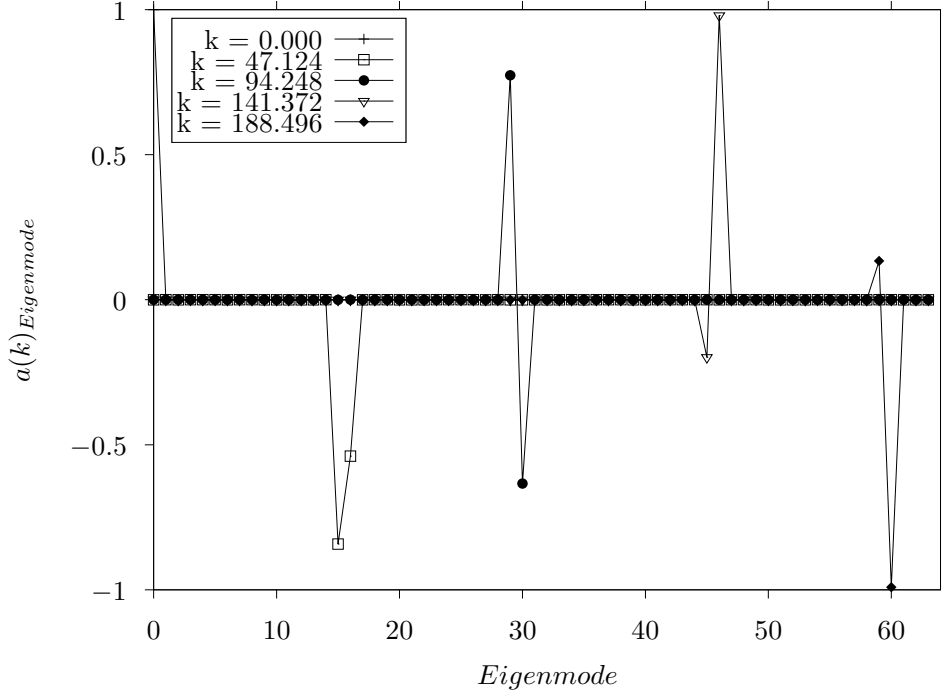
\begin{figure}
 \centering
 \resizebox{0.8\textwidth}{!}{\input{eigen2SinAmatrix1000.tex}}
 \caption{$\isdisc{a}(k)$ at different wavenumbers $k$ for uniform mesh.}\label{eigen2SinAmatrix1000}
\end{figure}
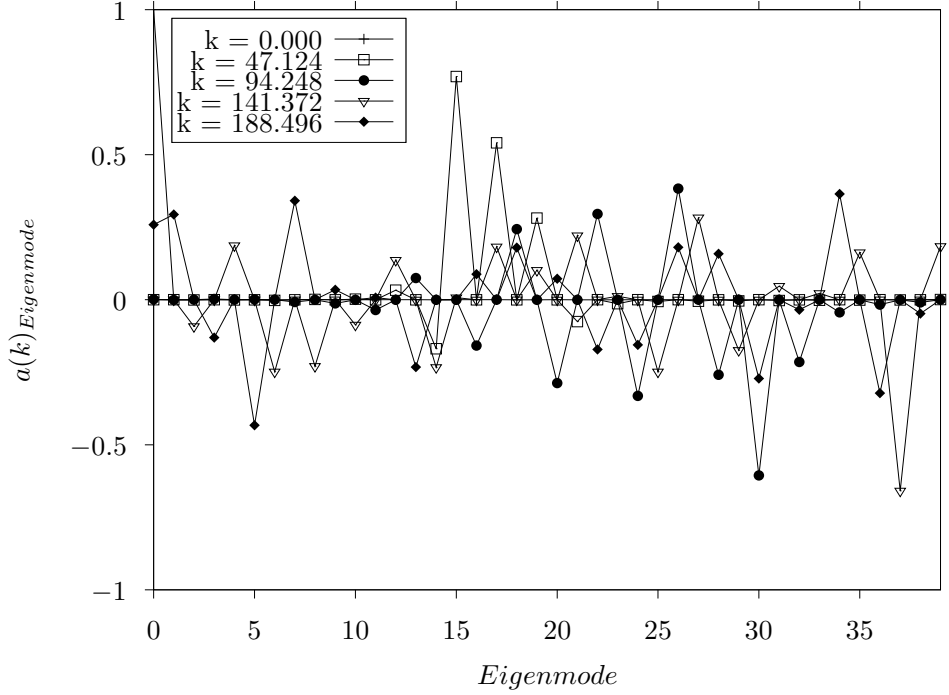
\begin{figure}
 \centering
 \resizebox{0.8\textwidth}{!}{\input{eigen2SinAmatrix1050.tex}}
 \caption{$\isdisc{a}(k)$ at different wavenumbers $k$ for mesh with 5\% stretching.}\label{eigen2SinAmatrix1050}
\end{figure}

As can be seen in figure \ref{eigen2SinAmatrix1000}, each frequency is represented by a single mode 
or by two consecutive in the worst case. This means each frequency is related to very few eigenmodes.
On the other hand, figure \ref{eigen2SinAmatrix1050} shows that each frequency is composed of a large 
amount of different eigenmodes; the higher the wavenumber, higher the number of eigenmodes that compose 
this frequency. Consequently, during the basis change using stretched meshes, the information contained
in matrix $\hat{\Gamma}$ is further dispersed. 

%% file: eigen2SinDiag.tex
% GNUPLOT: LaTeX picture with Postscript
\begingroup
  \makeatletter
  \providecommand\color[2][]{%
    \GenericError{(gnuplot) \space\space\space\@spaces}{%
      Package color not loaded in conjunction with
      terminal option `colourtext'%
    }{See the gnuplot documentation for explanation.%
    }{Either use 'blacktext' in gnuplot or load the package
      color.sty in LaTeX.}%
    \renewcommand\color[2][]{}%
  }%
  \providecommand\includegraphics[2][]{%
    \GenericError{(gnuplot) \space\space\space\@spaces}{%
      Package graphicx or graphics not loaded%
    }{See the gnuplot documentation for explanation.%
    }{The gnuplot epslatex terminal needs graphicx.sty or graphics.sty.}%
    \renewcommand\includegraphics[2][]{}%
  }%
  \providecommand\rotatebox[2]{#2}%
  \@ifundefined{ifGPcolor}{%
    \newif\ifGPcolor
    \GPcolorfalse
  }{}%
  \@ifundefined{ifGPblacktext}{%
    \newif\ifGPblacktext
    \GPblacktexttrue
  }{}%
  % define a \g@addto@macro without @ in the name:
  \let\gplgaddtomacro\g@addto@macro
  % define empty templates for all commands taking text:
  \gdef\gplbacktext{}%
  \gdef\gplfronttext{}%
  \makeatother
  \ifGPblacktext
    % no textcolor at all
    \def\colorrgb#1{}%
    \def\colorgray#1{}%
  \else
    % gray or color?
    \ifGPcolor
      \def\colorrgb#1{\color[rgb]{#1}}%
      \def\colorgray#1{\color[gray]{#1}}%
      \expandafter\def\csname LTw\endcsname{\color{white}}%
      \expandafter\def\csname LTb\endcsname{\color{black}}%
      \expandafter\def\csname LTa\endcsname{\color{black}}%
      \expandafter\def\csname LT0\endcsname{\color[rgb]{1,0,0}}%
      \expandafter\def\csname LT1\endcsname{\color[rgb]{0,1,0}}%
      \expandafter\def\csname LT2\endcsname{\color[rgb]{0,0,1}}%
      \expandafter\def\csname LT3\endcsname{\color[rgb]{1,0,1}}%
      \expandafter\def\csname LT4\endcsname{\color[rgb]{0,1,1}}%
      \expandafter\def\csname LT5\endcsname{\color[rgb]{1,1,0}}%
      \expandafter\def\csname LT6\endcsname{\color[rgb]{0,0,0}}%
      \expandafter\def\csname LT7\endcsname{\color[rgb]{1,0.3,0}}%
      \expandafter\def\csname LT8\endcsname{\color[rgb]{0.5,0.5,0.5}}%
    \else
      % gray
      \def\colorrgb#1{\color{black}}%
      \def\colorgray#1{\color[gray]{#1}}%
      \expandafter\def\csname LTw\endcsname{\color{white}}%
      \expandafter\def\csname LTb\endcsname{\color{black}}%
      \expandafter\def\csname LTa\endcsname{\color{black}}%
      \expandafter\def\csname LT0\endcsname{\color{black}}%
      \expandafter\def\csname LT1\endcsname{\color{black}}%
      \expandafter\def\csname LT2\endcsname{\color{black}}%
      \expandafter\def\csname LT3\endcsname{\color{black}}%
      \expandafter\def\csname LT4\endcsname{\color{black}}%
      \expandafter\def\csname LT5\endcsname{\color{black}}%
      \expandafter\def\csname LT6\endcsname{\color{black}}%
      \expandafter\def\csname LT7\endcsname{\color{black}}%
      \expandafter\def\csname LT8\endcsname{\color{black}}%
    \fi
  \fi
    \setlength{\unitlength}{0.0500bp}%
    \ifx\gptboxheight\undefined%
      \newlength{\gptboxheight}%
      \newlength{\gptboxwidth}%
      \newsavebox{\gptboxtext}%
    \fi%
    \setlength{\fboxrule}{0.5pt}%
    \setlength{\fboxsep}{1pt}%
\begin{picture}(7200.00,5040.00)%
    \gplgaddtomacro\gplbacktext{%
      \csname LTb\endcsname%
      \put(814,967){\makebox(0,0)[r]{\strut{}$-30$}}%
      \put(814,1368){\makebox(0,0)[r]{\strut{}$-20$}}%
      \put(814,1769){\makebox(0,0)[r]{\strut{}$-10$}}%
      \put(814,2170){\makebox(0,0)[r]{\strut{}$0$}}%
      \put(814,2571){\makebox(0,0)[r]{\strut{}$10$}}%
      \put(814,2972){\makebox(0,0)[r]{\strut{}$20$}}%
      \put(814,3373){\makebox(0,0)[r]{\strut{}$30$}}%
      \put(814,3773){\makebox(0,0)[r]{\strut{}$40$}}%
      \put(814,4174){\makebox(0,0)[r]{\strut{}$50$}}%
      \put(814,4575){\makebox(0,0)[r]{\strut{}$60$}}%
      \put(814,4976){\makebox(0,0)[r]{\strut{}$70$}}%
      \put(1009,484){\makebox(0,0){\strut{}$0$}}%
      \put(2450,484){\makebox(0,0){\strut{}$50$}}%
      \put(3891,484){\makebox(0,0){\strut{}$100$}}%
      \put(5332,484){\makebox(0,0){\strut{}$150$}}%
      \put(6772,484){\makebox(0,0){\strut{}$200$}}%
    }%
    \gplgaddtomacro\gplfronttext{%
      \csname LTb\endcsname%
      \put(176,2871){\rotatebox{-270}{\makebox(0,0){\strut{}$S(k,k) [m^{-1}]$}}}%
      \put(3906,154){\makebox(0,0){\strut{}$k [m^{-1}]$}}%
      \csname LTb\endcsname%
      \put(3874,2405){\makebox(0,0)[r]{\strut{}1.000}}%
      \put(3874,2255){\makebox(0,0)[r]{\strut{}1.005}}%
      \put(3874,2105){\makebox(0,0)[r]{\strut{}1.010}}%
      \put(3874,1955){\makebox(0,0)[r]{\strut{}1.015}}%
      \put(3874,1805){\makebox(0,0)[r]{\strut{}1.020}}%
      \put(3874,1655){\makebox(0,0)[r]{\strut{}1.025}}%
      \put(3874,1505){\makebox(0,0)[r]{\strut{}1.030}}%
      \put(3874,1355){\makebox(0,0)[r]{\strut{}1.035}}%
      \put(3874,1205){\makebox(0,0)[r]{\strut{}1.040}}%
      \put(3874,1055){\makebox(0,0)[r]{\strut{}1.045}}%
      \put(3874,905){\makebox(0,0)[r]{\strut{}1.050}}%
    }%
    \gplbacktext
    \put(0,0){\includegraphics{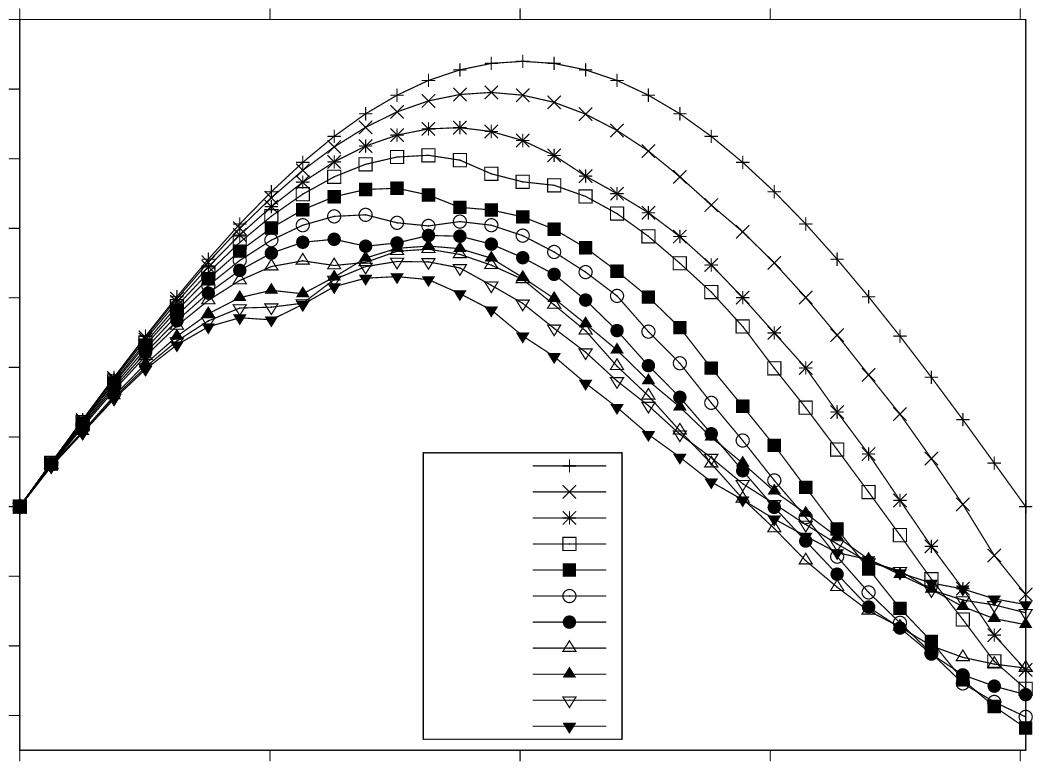}}%
    \gplfronttext
  \end{picture}%
\endgroup

%% file: eigen2SinLam.tex
% GNUPLOT: LaTeX picture with Postscript
\begingroup
  \makeatletter
  \providecommand\color[2][]{%
    \GenericError{(gnuplot) \space\space\space\@spaces}{%
      Package color not loaded in conjunction with
      terminal option `colourtext'%
    }{See the gnuplot documentation for explanation.%
    }{Either use 'blacktext' in gnuplot or load the package
      color.sty in LaTeX.}%
    \renewcommand\color[2][]{}%
  }%
  \providecommand\includegraphics[2][]{%
    \GenericError{(gnuplot) \space\space\space\@spaces}{%
      Package graphicx or graphics not loaded%
    }{See the gnuplot documentation for explanation.%
    }{The gnuplot epslatex terminal needs graphicx.sty or graphics.sty.}%
    \renewcommand\includegraphics[2][]{}%
  }%
  \providecommand\rotatebox[2]{#2}%
  \@ifundefined{ifGPcolor}{%
    \newif\ifGPcolor
    \GPcolorfalse
  }{}%
  \@ifundefined{ifGPblacktext}{%
    \newif\ifGPblacktext
    \GPblacktexttrue
  }{}%
  % define a \g@addto@macro without @ in the name:
  \let\gplgaddtomacro\g@addto@macro
  % define empty templates for all commands taking text:
  \gdef\gplbacktext{}%
  \gdef\gplfronttext{}%
  \makeatother
  \ifGPblacktext
    % no textcolor at all
    \def\colorrgb#1{}%
    \def\colorgray#1{}%
  \else
    % gray or color?
    \ifGPcolor
      \def\colorrgb#1{\color[rgb]{#1}}%
      \def\colorgray#1{\color[gray]{#1}}%
      \expandafter\def\csname LTw\endcsname{\color{white}}%
      \expandafter\def\csname LTb\endcsname{\color{black}}%
      \expandafter\def\csname LTa\endcsname{\color{black}}%
      \expandafter\def\csname LT0\endcsname{\color[rgb]{1,0,0}}%
      \expandafter\def\csname LT1\endcsname{\color[rgb]{0,1,0}}%
      \expandafter\def\csname LT2\endcsname{\color[rgb]{0,0,1}}%
      \expandafter\def\csname LT3\endcsname{\color[rgb]{1,0,1}}%
      \expandafter\def\csname LT4\endcsname{\color[rgb]{0,1,1}}%
      \expandafter\def\csname LT5\endcsname{\color[rgb]{1,1,0}}%
      \expandafter\def\csname LT6\endcsname{\color[rgb]{0,0,0}}%
      \expandafter\def\csname LT7\endcsname{\color[rgb]{1,0.3,0}}%
      \expandafter\def\csname LT8\endcsname{\color[rgb]{0.5,0.5,0.5}}%
    \else
      % gray
      \def\colorrgb#1{\color{black}}%
      \def\colorgray#1{\color[gray]{#1}}%
      \expandafter\def\csname LTw\endcsname{\color{white}}%
      \expandafter\def\csname LTb\endcsname{\color{black}}%
      \expandafter\def\csname LTa\endcsname{\color{black}}%
      \expandafter\def\csname LT0\endcsname{\color{black}}%
      \expandafter\def\csname LT1\endcsname{\color{black}}%
      \expandafter\def\csname LT2\endcsname{\color{black}}%
      \expandafter\def\csname LT3\endcsname{\color{black}}%
      \expandafter\def\csname LT4\endcsname{\color{black}}%
      \expandafter\def\csname LT5\endcsname{\color{black}}%
      \expandafter\def\csname LT6\endcsname{\color{black}}%
      \expandafter\def\csname LT7\endcsname{\color{black}}%
      \expandafter\def\csname LT8\endcsname{\color{black}}%
    \fi
  \fi
    \setlength{\unitlength}{0.0500bp}%
    \ifx\gptboxheight\undefined%
      \newlength{\gptboxheight}%
      \newlength{\gptboxwidth}%
      \newsavebox{\gptboxtext}%
    \fi%
    \setlength{\fboxrule}{0.5pt}%
    \setlength{\fboxsep}{1pt}%
\begin{picture}(7200.00,5040.00)%
    \gplgaddtomacro\gplbacktext{%
      \csname LTb\endcsname%
      \put(682,767){\makebox(0,0)[r]{\strut{}$0$}}%
      \put(682,1368){\makebox(0,0)[r]{\strut{}$10$}}%
      \put(682,1970){\makebox(0,0)[r]{\strut{}$20$}}%
      \put(682,2571){\makebox(0,0)[r]{\strut{}$30$}}%
      \put(682,3172){\makebox(0,0)[r]{\strut{}$40$}}%
      \put(682,3773){\makebox(0,0)[r]{\strut{}$50$}}%
      \put(682,4375){\makebox(0,0)[r]{\strut{}$60$}}%
      \put(682,4976){\makebox(0,0)[r]{\strut{}$70$}}%
      \put(877,484){\makebox(0,0){\strut{}$0$}}%
      \put(2351,484){\makebox(0,0){\strut{}$50$}}%
      \put(3824,484){\makebox(0,0){\strut{}$100$}}%
      \put(5298,484){\makebox(0,0){\strut{}$150$}}%
      \put(6772,484){\makebox(0,0){\strut{}$200$}}%
    }%
    \gplgaddtomacro\gplfronttext{%
      \csname LTb\endcsname%
      \put(176,2871){\rotatebox{-270}{\makebox(0,0){\strut{}$S_{rms}(k) [m^{-1}]$}}}%
      \put(3840,154){\makebox(0,0){\strut{}$k [m^{-1}]$}}%
      \csname LTb\endcsname%
      \put(3808,2405){\makebox(0,0)[r]{\strut{}1.000}}%
      \put(3808,2255){\makebox(0,0)[r]{\strut{}1.005}}%
      \put(3808,2105){\makebox(0,0)[r]{\strut{}1.010}}%
      \put(3808,1955){\makebox(0,0)[r]{\strut{}1.015}}%
      \put(3808,1805){\makebox(0,0)[r]{\strut{}1.020}}%
      \put(3808,1655){\makebox(0,0)[r]{\strut{}1.025}}%
      \put(3808,1505){\makebox(0,0)[r]{\strut{}1.030}}%
      \put(3808,1355){\makebox(0,0)[r]{\strut{}1.035}}%
      \put(3808,1205){\makebox(0,0)[r]{\strut{}1.040}}%
      \put(3808,1055){\makebox(0,0)[r]{\strut{}1.045}}%
      \put(3808,905){\makebox(0,0)[r]{\strut{}1.050}}%
    }%
    \gplbacktext
    \put(0,0){\includegraphics{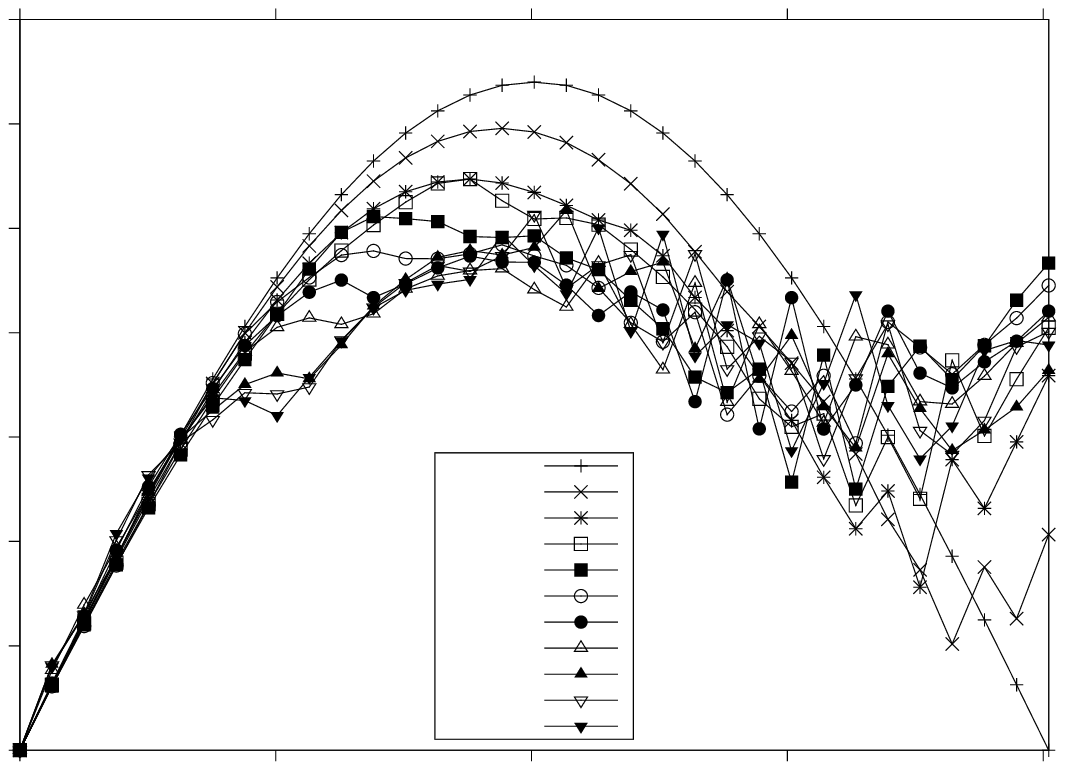}}%
    \gplfronttext
  \end{picture}%
\endgroup

%% file: aliasingMatrix1000.tex
% GNUPLOT: LaTeX picture with Postscript
\begingroup
  \makeatletter
  \providecommand\color[2][]{%
    \GenericError{(gnuplot) \space\space\space\@spaces}{%
      Package color not loaded in conjunction with
      terminal option `colourtext'%
    }{See the gnuplot documentation for explanation.%
    }{Either use 'blacktext' in gnuplot or load the package
      color.sty in LaTeX.}%
    \renewcommand\color[2][]{}%
  }%
  \providecommand\includegraphics[2][]{%
    \GenericError{(gnuplot) \space\space\space\@spaces}{%
      Package graphicx or graphics not loaded%
    }{See the gnuplot documentation for explanation.%
    }{The gnuplot epslatex terminal needs graphicx.sty or graphics.sty.}%
    \renewcommand\includegraphics[2][]{}%
  }%
  \providecommand\rotatebox[2]{#2}%
  \@ifundefined{ifGPcolor}{%
    \newif\ifGPcolor
    \GPcolorfalse
  }{}%
  \@ifundefined{ifGPblacktext}{%
    \newif\ifGPblacktext
    \GPblacktexttrue
  }{}%
  % define a \g@addto@macro without @ in the name:
  \let\gplgaddtomacro\g@addto@macro
  % define empty templates for all commands taking text:
  \gdef\gplbacktext{}%
  \gdef\gplfronttext{}%
  \makeatother
  \ifGPblacktext
    % no textcolor at all
    \def\colorrgb#1{}%
    \def\colorgray#1{}%
  \else
    % gray or color?
    \ifGPcolor
      \def\colorrgb#1{\color[rgb]{#1}}%
      \def\colorgray#1{\color[gray]{#1}}%
      \expandafter\def\csname LTw\endcsname{\color{white}}%
      \expandafter\def\csname LTb\endcsname{\color{black}}%
      \expandafter\def\csname LTa\endcsname{\color{black}}%
      \expandafter\def\csname LT0\endcsname{\color[rgb]{1,0,0}}%
      \expandafter\def\csname LT1\endcsname{\color[rgb]{0,1,0}}%
      \expandafter\def\csname LT2\endcsname{\color[rgb]{0,0,1}}%
      \expandafter\def\csname LT3\endcsname{\color[rgb]{1,0,1}}%
      \expandafter\def\csname LT4\endcsname{\color[rgb]{0,1,1}}%
      \expandafter\def\csname LT5\endcsname{\color[rgb]{1,1,0}}%
      \expandafter\def\csname LT6\endcsname{\color[rgb]{0,0,0}}%
      \expandafter\def\csname LT7\endcsname{\color[rgb]{1,0.3,0}}%
      \expandafter\def\csname LT8\endcsname{\color[rgb]{0.5,0.5,0.5}}%
    \else
      % gray
      \def\colorrgb#1{\color{black}}%
      \def\colorgray#1{\color[gray]{#1}}%
      \expandafter\def\csname LTw\endcsname{\color{white}}%
      \expandafter\def\csname LTb\endcsname{\color{black}}%
      \expandafter\def\csname LTa\endcsname{\color{black}}%
      \expandafter\def\csname LT0\endcsname{\color{black}}%
      \expandafter\def\csname LT1\endcsname{\color{black}}%
      \expandafter\def\csname LT2\endcsname{\color{black}}%
      \expandafter\def\csname LT3\endcsname{\color{black}}%
      \expandafter\def\csname LT4\endcsname{\color{black}}%
      \expandafter\def\csname LT5\endcsname{\color{black}}%
      \expandafter\def\csname LT6\endcsname{\color{black}}%
      \expandafter\def\csname LT7\endcsname{\color{black}}%
      \expandafter\def\csname LT8\endcsname{\color{black}}%
    \fi
  \fi
    \setlength{\unitlength}{0.0500bp}%
    \ifx\gptboxheight\undefined%
      \newlength{\gptboxheight}%
      \newlength{\gptboxwidth}%
      \newsavebox{\gptboxtext}%
    \fi%
    \setlength{\fboxrule}{0.5pt}%
    \setlength{\fboxsep}{1pt}%
\begin{picture}(7200.00,5040.00)%
    \gplgaddtomacro\gplbacktext{%
    }%
    \gplgaddtomacro\gplfronttext{%
      \csname LTb\endcsname%
      \put(4562,1331){\makebox(0,0)[r]{\strut{}Positive}}%
      \csname LTb\endcsname%
      \put(4562,1111){\makebox(0,0)[r]{\strut{}Negative}}%
      \csname LTb\endcsname%
      \put(1720,534){\makebox(0,0){\strut{}$0$}}%
      \put(2307,534){\makebox(0,0){\strut{}$10$}}%
      \put(2895,534){\makebox(0,0){\strut{}$20$}}%
      \put(3483,534){\makebox(0,0){\strut{}$30$}}%
      \put(4070,534){\makebox(0,0){\strut{}$40$}}%
      \put(4658,534){\makebox(0,0){\strut{}$50$}}%
      \put(5245,534){\makebox(0,0){\strut{}$60$}}%
      \put(3600,204){\makebox(0,0){\strut{}$Mode_{In}$}}%
      \put(1442,938){\makebox(0,0)[r]{\strut{}$0$}}%
      \put(1442,1467){\makebox(0,0)[r]{\strut{}$10$}}%
      \put(1442,1996){\makebox(0,0)[r]{\strut{}$20$}}%
      \put(1442,2525){\makebox(0,0)[r]{\strut{}$30$}}%
      \put(1442,3053){\makebox(0,0)[r]{\strut{}$40$}}%
      \put(1442,3582){\makebox(0,0)[r]{\strut{}$50$}}%
      \put(1442,4111){\makebox(0,0)[r]{\strut{}$60$}}%
      \put(1112,2630){\rotatebox{-270}{\makebox(0,0){\strut{}$Mode_{Out}$}}}%
      \csname LTb\endcsname%
      \put(4562,1331){\makebox(0,0)[r]{\strut{}Positive}}%
      \csname LTb\endcsname%
      \put(4562,1111){\makebox(0,0)[r]{\strut{}Negative}}%
      \csname LTb\endcsname%
      \put(1720,534){\makebox(0,0){\strut{}$0$}}%
      \put(2307,534){\makebox(0,0){\strut{}$10$}}%
      \put(2895,534){\makebox(0,0){\strut{}$20$}}%
      \put(3483,534){\makebox(0,0){\strut{}$30$}}%
      \put(4070,534){\makebox(0,0){\strut{}$40$}}%
      \put(4658,534){\makebox(0,0){\strut{}$50$}}%
      \put(5245,534){\makebox(0,0){\strut{}$60$}}%
      \put(3600,204){\makebox(0,0){\strut{}$Mode_{In}$}}%
      \put(1442,938){\makebox(0,0)[r]{\strut{}$0$}}%
      \put(1442,1467){\makebox(0,0)[r]{\strut{}$10$}}%
      \put(1442,1996){\makebox(0,0)[r]{\strut{}$20$}}%
      \put(1442,2525){\makebox(0,0)[r]{\strut{}$30$}}%
      \put(1442,3053){\makebox(0,0)[r]{\strut{}$40$}}%
      \put(1442,3582){\makebox(0,0)[r]{\strut{}$50$}}%
      \put(1442,4111){\makebox(0,0)[r]{\strut{}$60$}}%
      \put(1112,2630){\rotatebox{-270}{\makebox(0,0){\strut{}$Mode_{Out}$}}}%
      \put(6021,1043){\makebox(0,0)[l]{\strut{}$-60$}}%
      \put(6021,1572){\makebox(0,0)[l]{\strut{}$-40$}}%
      \put(6021,2101){\makebox(0,0)[l]{\strut{}$-20$}}%
      \put(6021,2630){\makebox(0,0)[l]{\strut{}$0$}}%
      \put(6021,3158){\makebox(0,0)[l]{\strut{}$20$}}%
      \put(6021,3687){\makebox(0,0)[l]{\strut{}$40$}}%
      \put(6021,4216){\makebox(0,0)[l]{\strut{}$60$}}%
      \put(6483,2630){\rotatebox{-270}{\makebox(0,0){\strut{}$S(k,k) [m^{-1}]$}}}%
    }%
    \gplbacktext
    \put(0,0){\includegraphics{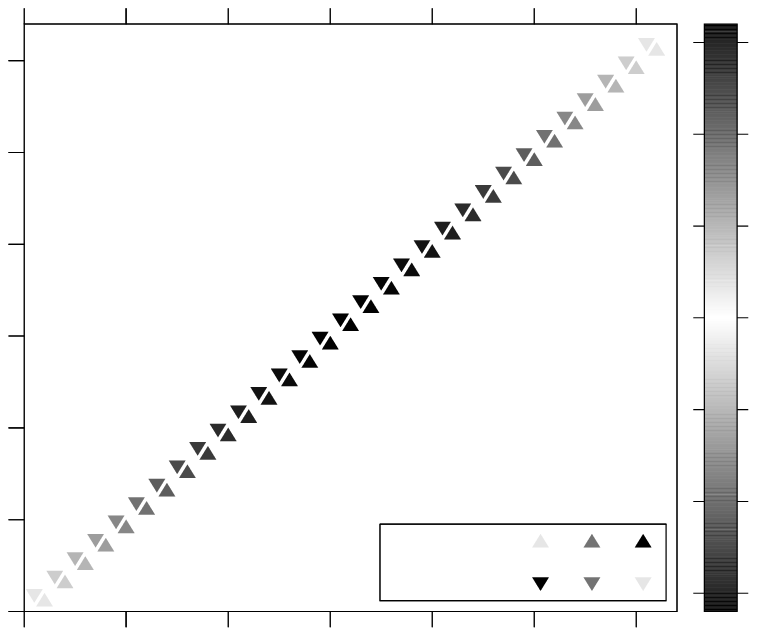}}%
    \gplfronttext
  \end{picture}%
\endgroup

%% file: aliasingMatrix1050.tex
% GNUPLOT: LaTeX picture with Postscript
\begingroup
  \makeatletter
  \providecommand\color[2][]{%
    \GenericError{(gnuplot) \space\space\space\@spaces}{%
      Package color not loaded in conjunction with
      terminal option `colourtext'%
    }{See the gnuplot documentation for explanation.%
    }{Either use 'blacktext' in gnuplot or load the package
      color.sty in LaTeX.}%
    \renewcommand\color[2][]{}%
  }%
  \providecommand\includegraphics[2][]{%
    \GenericError{(gnuplot) \space\space\space\@spaces}{%
      Package graphicx or graphics not loaded%
    }{See the gnuplot documentation for explanation.%
    }{The gnuplot epslatex terminal needs graphicx.sty or graphics.sty.}%
    \renewcommand\includegraphics[2][]{}%
  }%
  \providecommand\rotatebox[2]{#2}%
  \@ifundefined{ifGPcolor}{%
    \newif\ifGPcolor
    \GPcolorfalse
  }{}%
  \@ifundefined{ifGPblacktext}{%
    \newif\ifGPblacktext
    \GPblacktexttrue
  }{}%
  % define a \g@addto@macro without @ in the name:
  \let\gplgaddtomacro\g@addto@macro
  % define empty templates for all commands taking text:
  \gdef\gplbacktext{}%
  \gdef\gplfronttext{}%
  \makeatother
  \ifGPblacktext
    % no textcolor at all
    \def\colorrgb#1{}%
    \def\colorgray#1{}%
  \else
    % gray or color?
    \ifGPcolor
      \def\colorrgb#1{\color[rgb]{#1}}%
      \def\colorgray#1{\color[gray]{#1}}%
      \expandafter\def\csname LTw\endcsname{\color{white}}%
      \expandafter\def\csname LTb\endcsname{\color{black}}%
      \expandafter\def\csname LTa\endcsname{\color{black}}%
      \expandafter\def\csname LT0\endcsname{\color[rgb]{1,0,0}}%
      \expandafter\def\csname LT1\endcsname{\color[rgb]{0,1,0}}%
      \expandafter\def\csname LT2\endcsname{\color[rgb]{0,0,1}}%
      \expandafter\def\csname LT3\endcsname{\color[rgb]{1,0,1}}%
      \expandafter\def\csname LT4\endcsname{\color[rgb]{0,1,1}}%
      \expandafter\def\csname LT5\endcsname{\color[rgb]{1,1,0}}%
      \expandafter\def\csname LT6\endcsname{\color[rgb]{0,0,0}}%
      \expandafter\def\csname LT7\endcsname{\color[rgb]{1,0.3,0}}%
      \expandafter\def\csname LT8\endcsname{\color[rgb]{0.5,0.5,0.5}}%
    \else
      % gray
      \def\colorrgb#1{\color{black}}%
      \def\colorgray#1{\color[gray]{#1}}%
      \expandafter\def\csname LTw\endcsname{\color{white}}%
      \expandafter\def\csname LTb\endcsname{\color{black}}%
      \expandafter\def\csname LTa\endcsname{\color{black}}%
      \expandafter\def\csname LT0\endcsname{\color{black}}%
      \expandafter\def\csname LT1\endcsname{\color{black}}%
      \expandafter\def\csname LT2\endcsname{\color{black}}%
      \expandafter\def\csname LT3\endcsname{\color{black}}%
      \expandafter\def\csname LT4\endcsname{\color{black}}%
      \expandafter\def\csname LT5\endcsname{\color{black}}%
      \expandafter\def\csname LT6\endcsname{\color{black}}%
      \expandafter\def\csname LT7\endcsname{\color{black}}%
      \expandafter\def\csname LT8\endcsname{\color{black}}%
    \fi
  \fi
    \setlength{\unitlength}{0.0500bp}%
    \ifx\gptboxheight\undefined%
      \newlength{\gptboxheight}%
      \newlength{\gptboxwidth}%
      \newsavebox{\gptboxtext}%
    \fi%
    \setlength{\fboxrule}{0.5pt}%
    \setlength{\fboxsep}{1pt}%
\begin{picture}(7200.00,5040.00)%
    \gplgaddtomacro\gplbacktext{%
    }%
    \gplgaddtomacro\gplfronttext{%
      \csname LTb\endcsname%
      \put(4562,1331){\makebox(0,0)[r]{\strut{}Positive}}%
      \csname LTb\endcsname%
      \put(4562,1111){\makebox(0,0)[r]{\strut{}Negative}}%
      \csname LTb\endcsname%
      \put(1720,534){\makebox(0,0){\strut{}$0$}}%
      \put(2307,534){\makebox(0,0){\strut{}$10$}}%
      \put(2895,534){\makebox(0,0){\strut{}$20$}}%
      \put(3483,534){\makebox(0,0){\strut{}$30$}}%
      \put(4070,534){\makebox(0,0){\strut{}$40$}}%
      \put(4658,534){\makebox(0,0){\strut{}$50$}}%
      \put(5245,534){\makebox(0,0){\strut{}$60$}}%
      \put(3600,204){\makebox(0,0){\strut{}$Mode_{In}$}}%
      \put(1442,938){\makebox(0,0)[r]{\strut{}$0$}}%
      \put(1442,1467){\makebox(0,0)[r]{\strut{}$10$}}%
      \put(1442,1996){\makebox(0,0)[r]{\strut{}$20$}}%
      \put(1442,2525){\makebox(0,0)[r]{\strut{}$30$}}%
      \put(1442,3053){\makebox(0,0)[r]{\strut{}$40$}}%
      \put(1442,3582){\makebox(0,0)[r]{\strut{}$50$}}%
      \put(1442,4111){\makebox(0,0)[r]{\strut{}$60$}}%
      \put(1112,2630){\rotatebox{-270}{\makebox(0,0){\strut{}$Mode_{Out}$}}}%
      \csname LTb\endcsname%
      \put(4562,1331){\makebox(0,0)[r]{\strut{}Positive}}%
      \csname LTb\endcsname%
      \put(4562,1111){\makebox(0,0)[r]{\strut{}Negative}}%
      \csname LTb\endcsname%
      \put(1720,534){\makebox(0,0){\strut{}$0$}}%
      \put(2307,534){\makebox(0,0){\strut{}$10$}}%
      \put(2895,534){\makebox(0,0){\strut{}$20$}}%
      \put(3483,534){\makebox(0,0){\strut{}$30$}}%
      \put(4070,534){\makebox(0,0){\strut{}$40$}}%
      \put(4658,534){\makebox(0,0){\strut{}$50$}}%
      \put(5245,534){\makebox(0,0){\strut{}$60$}}%
      \put(3600,204){\makebox(0,0){\strut{}$Mode_{In}$}}%
      \put(1442,938){\makebox(0,0)[r]{\strut{}$0$}}%
      \put(1442,1467){\makebox(0,0)[r]{\strut{}$10$}}%
      \put(1442,1996){\makebox(0,0)[r]{\strut{}$20$}}%
      \put(1442,2525){\makebox(0,0)[r]{\strut{}$30$}}%
      \put(1442,3053){\makebox(0,0)[r]{\strut{}$40$}}%
      \put(1442,3582){\makebox(0,0)[r]{\strut{}$50$}}%
      \put(1442,4111){\makebox(0,0)[r]{\strut{}$60$}}%
      \put(1112,2630){\rotatebox{-270}{\makebox(0,0){\strut{}$Mode_{Out}$}}}%
      \put(6021,1043){\makebox(0,0)[l]{\strut{}$-60$}}%
      \put(6021,1572){\makebox(0,0)[l]{\strut{}$-40$}}%
      \put(6021,2101){\makebox(0,0)[l]{\strut{}$-20$}}%
      \put(6021,2630){\makebox(0,0)[l]{\strut{}$0$}}%
      \put(6021,3158){\makebox(0,0)[l]{\strut{}$20$}}%
      \put(6021,3687){\makebox(0,0)[l]{\strut{}$40$}}%
      \put(6021,4216){\makebox(0,0)[l]{\strut{}$60$}}%
      \put(6483,2630){\rotatebox{-270}{\makebox(0,0){\strut{}$S(k,k) [m^{-1}]$}}}%
    }%
    \gplbacktext
    \put(0,0){\includegraphics{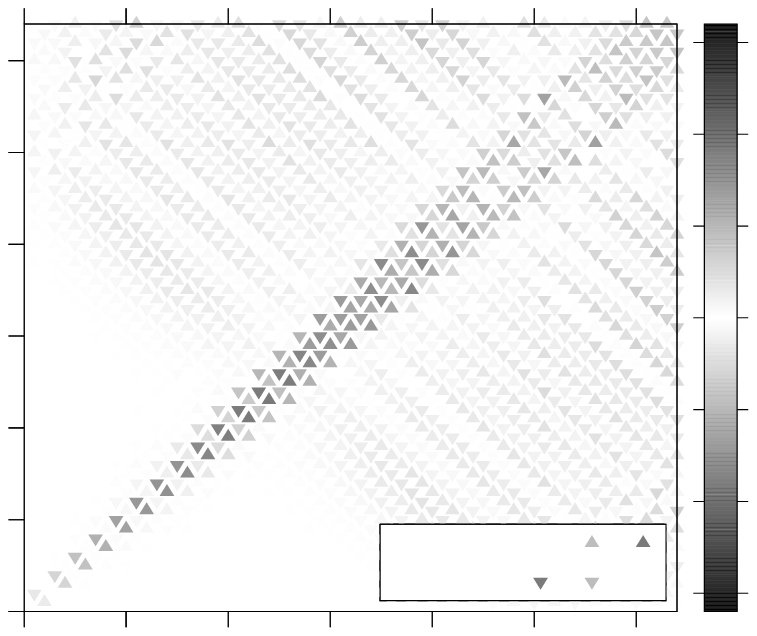}}%
    \gplfronttext
  \end{picture}%
\endgroup

%% file: eigen2eigenDiag.tex
% GNUPLOT: LaTeX picture with Postscript
\begingroup
  \makeatletter
  \providecommand\color[2][]{%
    \GenericError{(gnuplot) \space\space\space\@spaces}{%
      Package color not loaded in conjunction with
      terminal option `colourtext'%
    }{See the gnuplot documentation for explanation.%
    }{Either use 'blacktext' in gnuplot or load the package
      color.sty in LaTeX.}%
    \renewcommand\color[2][]{}%
  }%
  \providecommand\includegraphics[2][]{%
    \GenericError{(gnuplot) \space\space\space\@spaces}{%
      Package graphicx or graphics not loaded%
    }{See the gnuplot documentation for explanation.%
    }{The gnuplot epslatex terminal needs graphicx.sty or graphics.sty.}%
    \renewcommand\includegraphics[2][]{}%
  }%
  \providecommand\rotatebox[2]{#2}%
  \@ifundefined{ifGPcolor}{%
    \newif\ifGPcolor
    \GPcolorfalse
  }{}%
  \@ifundefined{ifGPblacktext}{%
    \newif\ifGPblacktext
    \GPblacktexttrue
  }{}%
  % define a \g@addto@macro without @ in the name:
  \let\gplgaddtomacro\g@addto@macro
  % define empty templates for all commands taking text:
  \gdef\gplbacktext{}%
  \gdef\gplfronttext{}%
  \makeatother
  \ifGPblacktext
    % no textcolor at all
    \def\colorrgb#1{}%
    \def\colorgray#1{}%
  \else
    % gray or color?
    \ifGPcolor
      \def\colorrgb#1{\color[rgb]{#1}}%
      \def\colorgray#1{\color[gray]{#1}}%
      \expandafter\def\csname LTw\endcsname{\color{white}}%
      \expandafter\def\csname LTb\endcsname{\color{black}}%
      \expandafter\def\csname LTa\endcsname{\color{black}}%
      \expandafter\def\csname LT0\endcsname{\color[rgb]{1,0,0}}%
      \expandafter\def\csname LT1\endcsname{\color[rgb]{0,1,0}}%
      \expandafter\def\csname LT2\endcsname{\color[rgb]{0,0,1}}%
      \expandafter\def\csname LT3\endcsname{\color[rgb]{1,0,1}}%
      \expandafter\def\csname LT4\endcsname{\color[rgb]{0,1,1}}%
      \expandafter\def\csname LT5\endcsname{\color[rgb]{1,1,0}}%
      \expandafter\def\csname LT6\endcsname{\color[rgb]{0,0,0}}%
      \expandafter\def\csname LT7\endcsname{\color[rgb]{1,0.3,0}}%
      \expandafter\def\csname LT8\endcsname{\color[rgb]{0.5,0.5,0.5}}%
    \else
      % gray
      \def\colorrgb#1{\color{black}}%
      \def\colorgray#1{\color[gray]{#1}}%
      \expandafter\def\csname LTw\endcsname{\color{white}}%
      \expandafter\def\csname LTb\endcsname{\color{black}}%
      \expandafter\def\csname LTa\endcsname{\color{black}}%
      \expandafter\def\csname LT0\endcsname{\color{black}}%
      \expandafter\def\csname LT1\endcsname{\color{black}}%
      \expandafter\def\csname LT2\endcsname{\color{black}}%
      \expandafter\def\csname LT3\endcsname{\color{black}}%
      \expandafter\def\csname LT4\endcsname{\color{black}}%
      \expandafter\def\csname LT5\endcsname{\color{black}}%
      \expandafter\def\csname LT6\endcsname{\color{black}}%
      \expandafter\def\csname LT7\endcsname{\color{black}}%
      \expandafter\def\csname LT8\endcsname{\color{black}}%
    \fi
  \fi
    \setlength{\unitlength}{0.0500bp}%
    \ifx\gptboxheight\undefined%
      \newlength{\gptboxheight}%
      \newlength{\gptboxwidth}%
      \newsavebox{\gptboxtext}%
    \fi%
    \setlength{\fboxrule}{0.5pt}%
    \setlength{\fboxsep}{1pt}%
\begin{picture}(7200.00,5040.00)%
    \gplgaddtomacro\gplbacktext{%
      \csname LTb\endcsname%
      \put(682,767){\makebox(0,0)[r]{\strut{}$0$}}%
      \put(682,1368){\makebox(0,0)[r]{\strut{}$10$}}%
      \put(682,1970){\makebox(0,0)[r]{\strut{}$20$}}%
      \put(682,2571){\makebox(0,0)[r]{\strut{}$30$}}%
      \put(682,3172){\makebox(0,0)[r]{\strut{}$40$}}%
      \put(682,3773){\makebox(0,0)[r]{\strut{}$50$}}%
      \put(682,4375){\makebox(0,0)[r]{\strut{}$60$}}%
      \put(682,4976){\makebox(0,0)[r]{\strut{}$70$}}%
      \put(877,484){\makebox(0,0){\strut{}$0$}}%
      \put(1803,484){\makebox(0,0){\strut{}$20$}}%
      \put(2729,484){\makebox(0,0){\strut{}$40$}}%
      \put(3655,484){\makebox(0,0){\strut{}$60$}}%
      \put(4581,484){\makebox(0,0){\strut{}$80$}}%
      \put(5507,484){\makebox(0,0){\strut{}$100$}}%
      \put(6433,484){\makebox(0,0){\strut{}$120$}}%
    }%
    \gplgaddtomacro\gplfronttext{%
      \csname LTb\endcsname%
      \put(176,2871){\rotatebox{-270}{\makebox(0,0){\strut{}$\hat{\gamma}(\lambda,\lambda) [m^{-1}]$}}}%
      \put(3840,154){\makebox(0,0){\strut{}$\lambda [m^{-1}]$}}%
      \csname LTb\endcsname%
      \put(1549,4838){\makebox(0,0)[r]{\strut{}1.000}}%
      \put(1549,4688){\makebox(0,0)[r]{\strut{}1.005}}%
      \put(1549,4538){\makebox(0,0)[r]{\strut{}1.010}}%
      \put(1549,4388){\makebox(0,0)[r]{\strut{}1.015}}%
      \put(1549,4238){\makebox(0,0)[r]{\strut{}1.020}}%
      \put(1549,4088){\makebox(0,0)[r]{\strut{}1.025}}%
      \put(1549,3938){\makebox(0,0)[r]{\strut{}1.030}}%
      \put(1549,3788){\makebox(0,0)[r]{\strut{}1.035}}%
      \put(1549,3638){\makebox(0,0)[r]{\strut{}1.040}}%
      \put(1549,3488){\makebox(0,0)[r]{\strut{}1.045}}%
      \put(1549,3338){\makebox(0,0)[r]{\strut{}1.050}}%
    }%
    \gplbacktext
    \put(0,0){\includegraphics{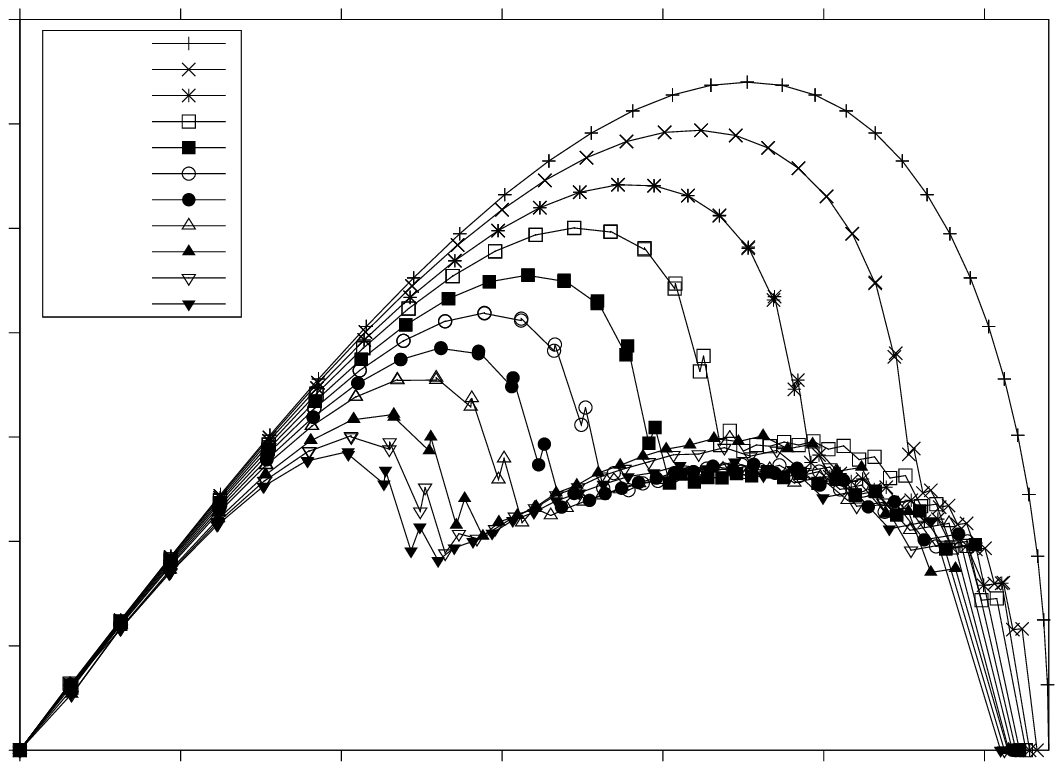}}%
    \gplfronttext
  \end{picture}%
\endgroup

%% file: eigen2eigenLam.tex
% GNUPLOT: LaTeX picture with Postscript
\begingroup
  \makeatletter
  \providecommand\color[2][]{%
    \GenericError{(gnuplot) \space\space\space\@spaces}{%
      Package color not loaded in conjunction with
      terminal option `colourtext'%
    }{See the gnuplot documentation for explanation.%
    }{Either use 'blacktext' in gnuplot or load the package
      color.sty in LaTeX.}%
    \renewcommand\color[2][]{}%
  }%
  \providecommand\includegraphics[2][]{%
    \GenericError{(gnuplot) \space\space\space\@spaces}{%
      Package graphicx or graphics not loaded%
    }{See the gnuplot documentation for explanation.%
    }{The gnuplot epslatex terminal needs graphicx.sty or graphics.sty.}%
    \renewcommand\includegraphics[2][]{}%
  }%
  \providecommand\rotatebox[2]{#2}%
  \@ifundefined{ifGPcolor}{%
    \newif\ifGPcolor
    \GPcolorfalse
  }{}%
  \@ifundefined{ifGPblacktext}{%
    \newif\ifGPblacktext
    \GPblacktexttrue
  }{}%
  % define a \g@addto@macro without @ in the name:
  \let\gplgaddtomacro\g@addto@macro
  % define empty templates for all commands taking text:
  \gdef\gplbacktext{}%
  \gdef\gplfronttext{}%
  \makeatother
  \ifGPblacktext
    % no textcolor at all
    \def\colorrgb#1{}%
    \def\colorgray#1{}%
  \else
    % gray or color?
    \ifGPcolor
      \def\colorrgb#1{\color[rgb]{#1}}%
      \def\colorgray#1{\color[gray]{#1}}%
      \expandafter\def\csname LTw\endcsname{\color{white}}%
      \expandafter\def\csname LTb\endcsname{\color{black}}%
      \expandafter\def\csname LTa\endcsname{\color{black}}%
      \expandafter\def\csname LT0\endcsname{\color[rgb]{1,0,0}}%
      \expandafter\def\csname LT1\endcsname{\color[rgb]{0,1,0}}%
      \expandafter\def\csname LT2\endcsname{\color[rgb]{0,0,1}}%
      \expandafter\def\csname LT3\endcsname{\color[rgb]{1,0,1}}%
      \expandafter\def\csname LT4\endcsname{\color[rgb]{0,1,1}}%
      \expandafter\def\csname LT5\endcsname{\color[rgb]{1,1,0}}%
      \expandafter\def\csname LT6\endcsname{\color[rgb]{0,0,0}}%
      \expandafter\def\csname LT7\endcsname{\color[rgb]{1,0.3,0}}%
      \expandafter\def\csname LT8\endcsname{\color[rgb]{0.5,0.5,0.5}}%
    \else
      % gray
      \def\colorrgb#1{\color{black}}%
      \def\colorgray#1{\color[gray]{#1}}%
      \expandafter\def\csname LTw\endcsname{\color{white}}%
      \expandafter\def\csname LTb\endcsname{\color{black}}%
      \expandafter\def\csname LTa\endcsname{\color{black}}%
      \expandafter\def\csname LT0\endcsname{\color{black}}%
      \expandafter\def\csname LT1\endcsname{\color{black}}%
      \expandafter\def\csname LT2\endcsname{\color{black}}%
      \expandafter\def\csname LT3\endcsname{\color{black}}%
      \expandafter\def\csname LT4\endcsname{\color{black}}%
      \expandafter\def\csname LT5\endcsname{\color{black}}%
      \expandafter\def\csname LT6\endcsname{\color{black}}%
      \expandafter\def\csname LT7\endcsname{\color{black}}%
      \expandafter\def\csname LT8\endcsname{\color{black}}%
    \fi
  \fi
    \setlength{\unitlength}{0.0500bp}%
    \ifx\gptboxheight\undefined%
      \newlength{\gptboxheight}%
      \newlength{\gptboxwidth}%
      \newsavebox{\gptboxtext}%
    \fi%
    \setlength{\fboxrule}{0.5pt}%
    \setlength{\fboxsep}{1pt}%
\begin{picture}(7200.00,5040.00)%
    \gplgaddtomacro\gplbacktext{%
      \csname LTb\endcsname%
      \put(682,767){\makebox(0,0)[r]{\strut{}$0$}}%
      \put(682,1368){\makebox(0,0)[r]{\strut{}$10$}}%
      \put(682,1970){\makebox(0,0)[r]{\strut{}$20$}}%
      \put(682,2571){\makebox(0,0)[r]{\strut{}$30$}}%
      \put(682,3172){\makebox(0,0)[r]{\strut{}$40$}}%
      \put(682,3773){\makebox(0,0)[r]{\strut{}$50$}}%
      \put(682,4375){\makebox(0,0)[r]{\strut{}$60$}}%
      \put(682,4976){\makebox(0,0)[r]{\strut{}$70$}}%
      \put(877,484){\makebox(0,0){\strut{}$0$}}%
      \put(1803,484){\makebox(0,0){\strut{}$20$}}%
      \put(2729,484){\makebox(0,0){\strut{}$40$}}%
      \put(3655,484){\makebox(0,0){\strut{}$60$}}%
      \put(4581,484){\makebox(0,0){\strut{}$80$}}%
      \put(5507,484){\makebox(0,0){\strut{}$100$}}%
      \put(6433,484){\makebox(0,0){\strut{}$120$}}%
    }%
    \gplgaddtomacro\gplfronttext{%
      \csname LTb\endcsname%
      \put(176,2871){\rotatebox{-270}{\makebox(0,0){\strut{}$\hat{\lambda}(\lambda)  [m^{-1}]$}}}%
      \put(3840,154){\makebox(0,0){\strut{}$\lambda [m^{-1}]$}}%
      \csname LTb\endcsname%
      \put(1549,4838){\makebox(0,0)[r]{\strut{}1.000}}%
      \put(1549,4688){\makebox(0,0)[r]{\strut{}1.005}}%
      \put(1549,4538){\makebox(0,0)[r]{\strut{}1.010}}%
      \put(1549,4388){\makebox(0,0)[r]{\strut{}1.015}}%
      \put(1549,4238){\makebox(0,0)[r]{\strut{}1.020}}%
      \put(1549,4088){\makebox(0,0)[r]{\strut{}1.025}}%
      \put(1549,3938){\makebox(0,0)[r]{\strut{}1.030}}%
      \put(1549,3788){\makebox(0,0)[r]{\strut{}1.035}}%
      \put(1549,3638){\makebox(0,0)[r]{\strut{}1.040}}%
      \put(1549,3488){\makebox(0,0)[r]{\strut{}1.045}}%
      \put(1549,3338){\makebox(0,0)[r]{\strut{}1.050}}%
    }%
    \gplbacktext
    \put(0,0){\includegraphics{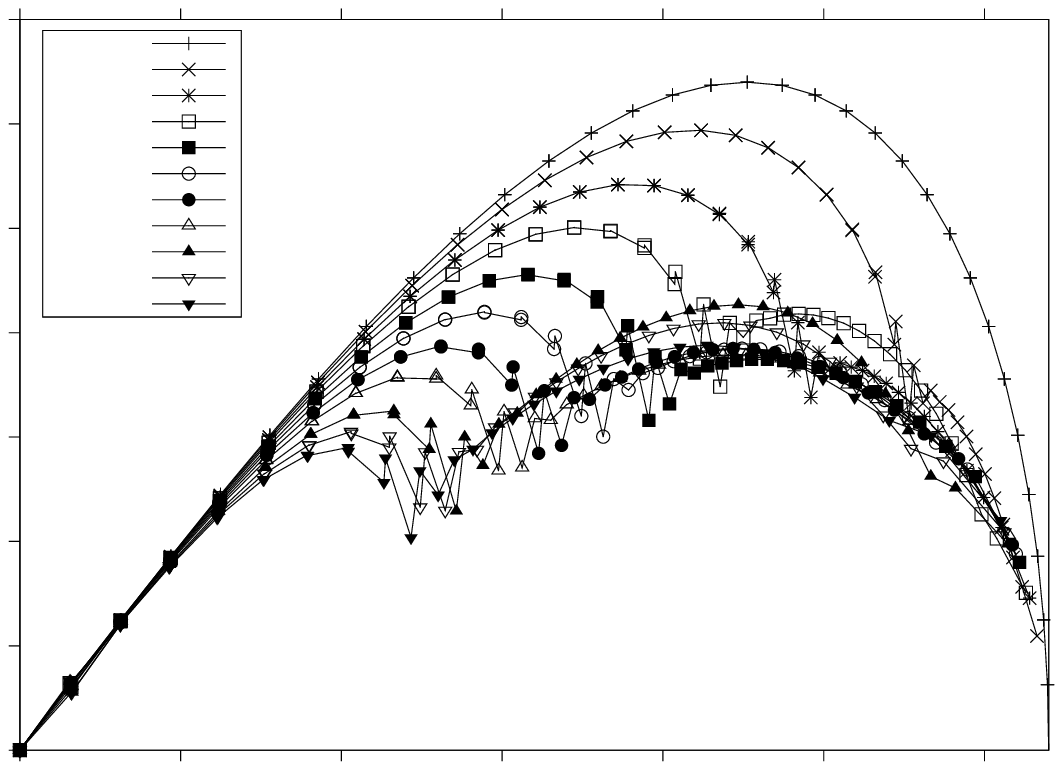}}%
    \gplfronttext
  \end{picture}%
\endgroup

%% file: eigen2SinAmatrix1000.tex
% GNUPLOT: LaTeX picture with Postscript
\begingroup
  \makeatletter
  \providecommand\color[2][]{%
    \GenericError{(gnuplot) \space\space\space\@spaces}{%
      Package color not loaded in conjunction with
      terminal option `colourtext'%
    }{See the gnuplot documentation for explanation.%
    }{Either use 'blacktext' in gnuplot or load the package
      color.sty in LaTeX.}%
    \renewcommand\color[2][]{}%
  }%
  \providecommand\includegraphics[2][]{%
    \GenericError{(gnuplot) \space\space\space\@spaces}{%
      Package graphicx or graphics not loaded%
    }{See the gnuplot documentation for explanation.%
    }{The gnuplot epslatex terminal needs graphicx.sty or graphics.sty.}%
    \renewcommand\includegraphics[2][]{}%
  }%
  \providecommand\rotatebox[2]{#2}%
  \@ifundefined{ifGPcolor}{%
    \newif\ifGPcolor
    \GPcolorfalse
  }{}%
  \@ifundefined{ifGPblacktext}{%
    \newif\ifGPblacktext
    \GPblacktexttrue
  }{}%
  % define a \g@addto@macro without @ in the name:
  \let\gplgaddtomacro\g@addto@macro
  % define empty templates for all commands taking text:
  \gdef\gplbacktext{}%
  \gdef\gplfronttext{}%
  \makeatother
  \ifGPblacktext
    % no textcolor at all
    \def\colorrgb#1{}%
    \def\colorgray#1{}%
  \else
    % gray or color?
    \ifGPcolor
      \def\colorrgb#1{\color[rgb]{#1}}%
      \def\colorgray#1{\color[gray]{#1}}%
      \expandafter\def\csname LTw\endcsname{\color{white}}%
      \expandafter\def\csname LTb\endcsname{\color{black}}%
      \expandafter\def\csname LTa\endcsname{\color{black}}%
      \expandafter\def\csname LT0\endcsname{\color[rgb]{1,0,0}}%
      \expandafter\def\csname LT1\endcsname{\color[rgb]{0,1,0}}%
      \expandafter\def\csname LT2\endcsname{\color[rgb]{0,0,1}}%
      \expandafter\def\csname LT3\endcsname{\color[rgb]{1,0,1}}%
      \expandafter\def\csname LT4\endcsname{\color[rgb]{0,1,1}}%
      \expandafter\def\csname LT5\endcsname{\color[rgb]{1,1,0}}%
      \expandafter\def\csname LT6\endcsname{\color[rgb]{0,0,0}}%
      \expandafter\def\csname LT7\endcsname{\color[rgb]{1,0.3,0}}%
      \expandafter\def\csname LT8\endcsname{\color[rgb]{0.5,0.5,0.5}}%
    \else
      % gray
      \def\colorrgb#1{\color{black}}%
      \def\colorgray#1{\color[gray]{#1}}%
      \expandafter\def\csname LTw\endcsname{\color{white}}%
      \expandafter\def\csname LTb\endcsname{\color{black}}%
      \expandafter\def\csname LTa\endcsname{\color{black}}%
      \expandafter\def\csname LT0\endcsname{\color{black}}%
      \expandafter\def\csname LT1\endcsname{\color{black}}%
      \expandafter\def\csname LT2\endcsname{\color{black}}%
      \expandafter\def\csname LT3\endcsname{\color{black}}%
      \expandafter\def\csname LT4\endcsname{\color{black}}%
      \expandafter\def\csname LT5\endcsname{\color{black}}%
      \expandafter\def\csname LT6\endcsname{\color{black}}%
      \expandafter\def\csname LT7\endcsname{\color{black}}%
      \expandafter\def\csname LT8\endcsname{\color{black}}%
    \fi
  \fi
    \setlength{\unitlength}{0.0500bp}%
    \ifx\gptboxheight\undefined%
      \newlength{\gptboxheight}%
      \newlength{\gptboxwidth}%
      \newsavebox{\gptboxtext}%
    \fi%
    \setlength{\fboxrule}{0.5pt}%
    \setlength{\fboxsep}{1pt}%
\begin{picture}(7200.00,5040.00)%
    \gplgaddtomacro\gplbacktext{%
      \csname LTb\endcsname%
      \put(946,767){\makebox(0,0)[r]{\strut{}$-1$}}%
      \put(946,1819){\makebox(0,0)[r]{\strut{}$-0.5$}}%
      \put(946,2872){\makebox(0,0)[r]{\strut{}$0$}}%
      \put(946,3924){\makebox(0,0)[r]{\strut{}$0.5$}}%
      \put(946,4976){\makebox(0,0)[r]{\strut{}$1$}}%
      \put(1141,484){\makebox(0,0){\strut{}$0$}}%
      \put(2026,484){\makebox(0,0){\strut{}$10$}}%
      \put(2910,484){\makebox(0,0){\strut{}$20$}}%
      \put(3795,484){\makebox(0,0){\strut{}$30$}}%
      \put(4680,484){\makebox(0,0){\strut{}$40$}}%
      \put(5564,484){\makebox(0,0){\strut{}$50$}}%
      \put(6449,484){\makebox(0,0){\strut{}$60$}}%
    }%
    \gplgaddtomacro\gplfronttext{%
      \csname LTb\endcsname%
      \put(176,2871){\rotatebox{-270}{\makebox(0,0){\strut{}$a(k)_{Eigenmode}$}}}%
      \put(3972,154){\makebox(0,0){\strut{}$Eigenmode$}}%
      \csname LTb\endcsname%
      \put(2353,4763){\makebox(0,0)[r]{\strut{}k = 0.000}}%
      \put(2353,4613){\makebox(0,0)[r]{\strut{}k = 47.124}}%
      \put(2353,4463){\makebox(0,0)[r]{\strut{}k = 94.248}}%
      \put(2353,4313){\makebox(0,0)[r]{\strut{}k = 141.372}}%
      \put(2353,4163){\makebox(0,0)[r]{\strut{}k = 188.496}}%
    }%
    \gplbacktext
    \put(0,0){\includegraphics{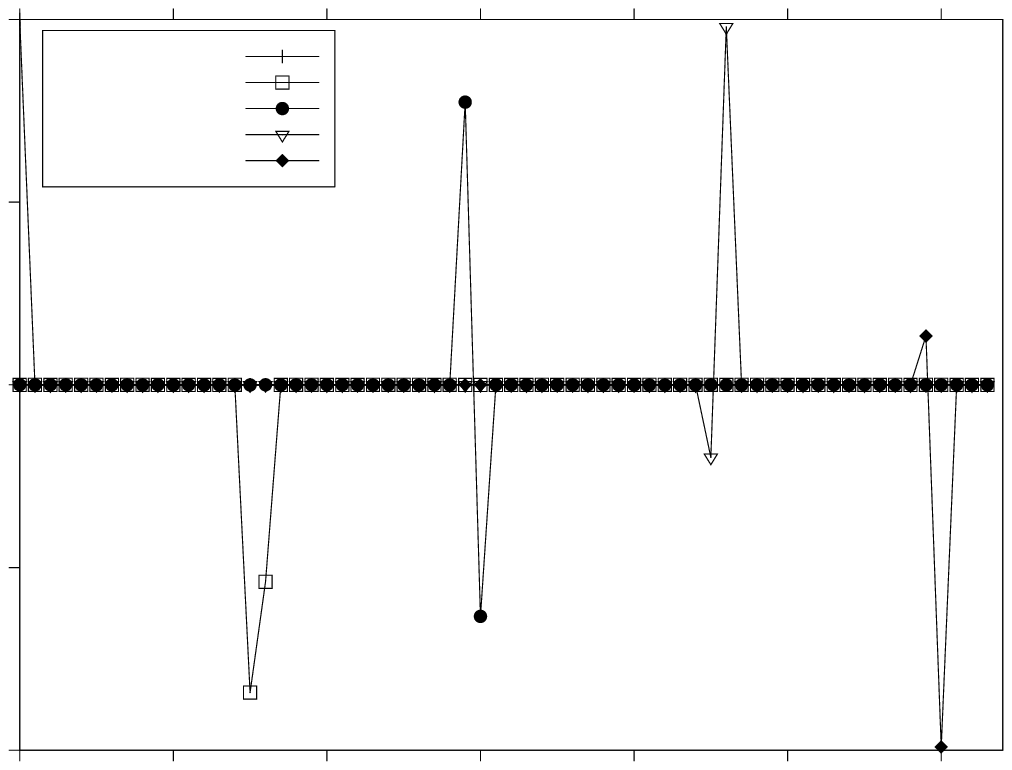}}%
    \gplfronttext
  \end{picture}%
\endgroup

%% file: eigen2SinAmatrix1050.tex
% GNUPLOT: LaTeX picture with Postscript
\begingroup
  \makeatletter
  \providecommand\color[2][]{%
    \GenericError{(gnuplot) \space\space\space\@spaces}{%
      Package color not loaded in conjunction with
      terminal option `colourtext'%
    }{See the gnuplot documentation for explanation.%
    }{Either use 'blacktext' in gnuplot or load the package
      color.sty in LaTeX.}%
    \renewcommand\color[2][]{}%
  }%
  \providecommand\includegraphics[2][]{%
    \GenericError{(gnuplot) \space\space\space\@spaces}{%
      Package graphicx or graphics not loaded%
    }{See the gnuplot documentation for explanation.%
    }{The gnuplot epslatex terminal needs graphicx.sty or graphics.sty.}%
    \renewcommand\includegraphics[2][]{}%
  }%
  \providecommand\rotatebox[2]{#2}%
  \@ifundefined{ifGPcolor}{%
    \newif\ifGPcolor
    \GPcolorfalse
  }{}%
  \@ifundefined{ifGPblacktext}{%
    \newif\ifGPblacktext
    \GPblacktexttrue
  }{}%
  % define a \g@addto@macro without @ in the name:
  \let\gplgaddtomacro\g@addto@macro
  % define empty templates for all commands taking text:
  \gdef\gplbacktext{}%
  \gdef\gplfronttext{}%
  \makeatother
  \ifGPblacktext
    % no textcolor at all
    \def\colorrgb#1{}%
    \def\colorgray#1{}%
  \else
    % gray or color?
    \ifGPcolor
      \def\colorrgb#1{\color[rgb]{#1}}%
      \def\colorgray#1{\color[gray]{#1}}%
      \expandafter\def\csname LTw\endcsname{\color{white}}%
      \expandafter\def\csname LTb\endcsname{\color{black}}%
      \expandafter\def\csname LTa\endcsname{\color{black}}%
      \expandafter\def\csname LT0\endcsname{\color[rgb]{1,0,0}}%
      \expandafter\def\csname LT1\endcsname{\color[rgb]{0,1,0}}%
      \expandafter\def\csname LT2\endcsname{\color[rgb]{0,0,1}}%
      \expandafter\def\csname LT3\endcsname{\color[rgb]{1,0,1}}%
      \expandafter\def\csname LT4\endcsname{\color[rgb]{0,1,1}}%
      \expandafter\def\csname LT5\endcsname{\color[rgb]{1,1,0}}%
      \expandafter\def\csname LT6\endcsname{\color[rgb]{0,0,0}}%
      \expandafter\def\csname LT7\endcsname{\color[rgb]{1,0.3,0}}%
      \expandafter\def\csname LT8\endcsname{\color[rgb]{0.5,0.5,0.5}}%
    \else
      % gray
      \def\colorrgb#1{\color{black}}%
      \def\colorgray#1{\color[gray]{#1}}%
      \expandafter\def\csname LTw\endcsname{\color{white}}%
      \expandafter\def\csname LTb\endcsname{\color{black}}%
      \expandafter\def\csname LTa\endcsname{\color{black}}%
      \expandafter\def\csname LT0\endcsname{\color{black}}%
      \expandafter\def\csname LT1\endcsname{\color{black}}%
      \expandafter\def\csname LT2\endcsname{\color{black}}%
      \expandafter\def\csname LT3\endcsname{\color{black}}%
      \expandafter\def\csname LT4\endcsname{\color{black}}%
      \expandafter\def\csname LT5\endcsname{\color{black}}%
      \expandafter\def\csname LT6\endcsname{\color{black}}%
      \expandafter\def\csname LT7\endcsname{\color{black}}%
      \expandafter\def\csname LT8\endcsname{\color{black}}%
    \fi
  \fi
    \setlength{\unitlength}{0.0500bp}%
    \ifx\gptboxheight\undefined%
      \newlength{\gptboxheight}%
      \newlength{\gptboxwidth}%
      \newsavebox{\gptboxtext}%
    \fi%
    \setlength{\fboxrule}{0.5pt}%
    \setlength{\fboxsep}{1pt}%
\begin{picture}(7200.00,5040.00)%
    \gplgaddtomacro\gplbacktext{%
      \csname LTb\endcsname%
      \put(946,767){\makebox(0,0)[r]{\strut{}$-1$}}%
      \put(946,1819){\makebox(0,0)[r]{\strut{}$-0.5$}}%
      \put(946,2872){\makebox(0,0)[r]{\strut{}$0$}}%
      \put(946,3924){\makebox(0,0)[r]{\strut{}$0.5$}}%
      \put(946,4976){\makebox(0,0)[r]{\strut{}$1$}}%
      \put(1141,484){\makebox(0,0){\strut{}$0$}}%
      \put(1867,484){\makebox(0,0){\strut{}$5$}}%
      \put(2593,484){\makebox(0,0){\strut{}$10$}}%
      \put(3319,484){\makebox(0,0){\strut{}$15$}}%
      \put(4045,484){\makebox(0,0){\strut{}$20$}}%
      \put(4770,484){\makebox(0,0){\strut{}$25$}}%
      \put(5496,484){\makebox(0,0){\strut{}$30$}}%
      \put(6222,484){\makebox(0,0){\strut{}$35$}}%
    }%
    \gplgaddtomacro\gplfronttext{%
      \csname LTb\endcsname%
      \put(176,2871){\rotatebox{-270}{\makebox(0,0){\strut{}$a(k)_{Eigenmode}$}}}%
      \put(3972,154){\makebox(0,0){\strut{}$Eigenmode$}}%
      \csname LTb\endcsname%
      \put(2353,4763){\makebox(0,0)[r]{\strut{}k = 0.000}}%
      \put(2353,4613){\makebox(0,0)[r]{\strut{}k = 47.124}}%
      \put(2353,4463){\makebox(0,0)[r]{\strut{}k = 94.248}}%
      \put(2353,4313){\makebox(0,0)[r]{\strut{}k = 141.372}}%
      \put(2353,4163){\makebox(0,0)[r]{\strut{}k = 188.496}}%
    }%
    \gplbacktext
    \put(0,0){\includegraphics{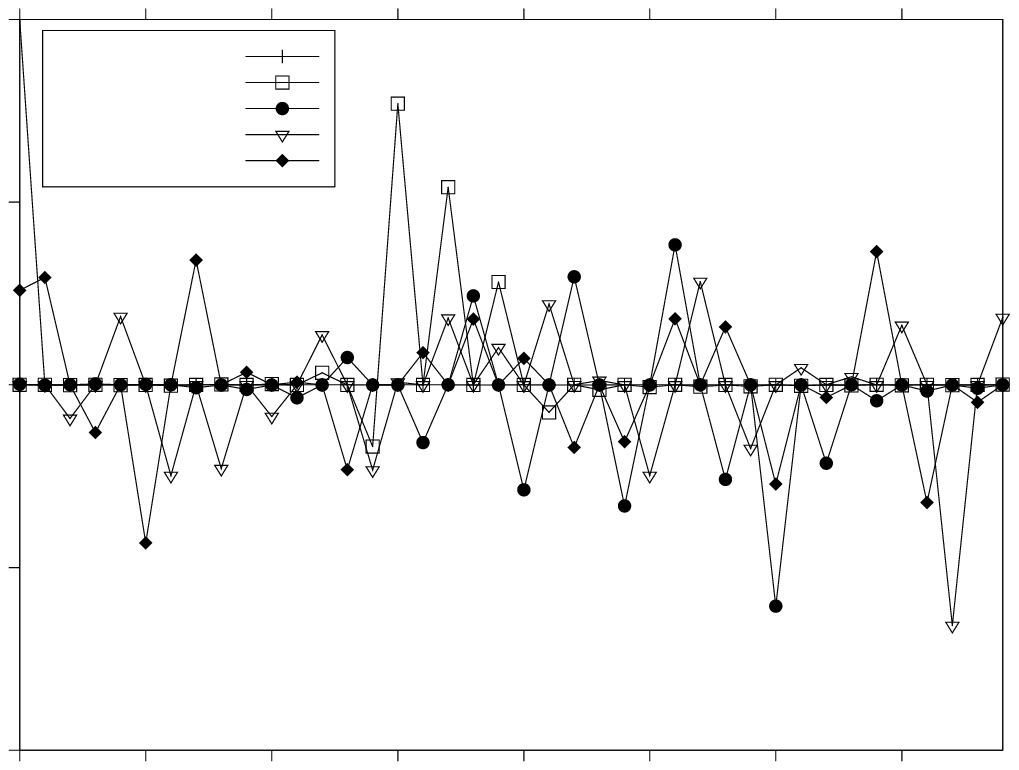}}%
    \gplfronttext
  \end{picture}%
\endgroup

%% file: 8.Conclusions.tex
We have developed an alternative methodology to compute the dispersion 
error not restricted to uniform meshes.
Due to the fact that the presented method does not require mesh 
uniformity, or more generaly a structured distribution of nodes, it has been tested 
using uniform and non-uniform structured meshes.
The obtained results in uniform mesh are the same obtained by Tam \cite{Tam1993}
and Lele \cite{Lele1992}; thus, the usual approach collapse on the current method
when uniform meshes are used.
The obtained results indicate that stretched meshes, even with small values around 
3-5\% degrade the solution significantly. Thus, the usage of smaller values is highly
recommended, as pointed by Shur \cite{Shur2005} or Bogey \cite{Bogey2004} by means of the
range of validity of their numerical results.
It also has been computed how non-uniform stretched meshes induce projection on 
other frequencies. This implies that the frequency of the approximate derivative 
is composed by several frequencies rather than a unique mode. Additionally,
it has been observed that the projection onto other frequencies performs through
two mechanisms in stretched meshes:
\begin{itemize}
 \item First, during the derivation process if the set of eigenvectors are not
 the eigenvectors of the discrete convective operator.
 \item Second, when projecting into the space of sinusoids and dispersing the
 information of the different eigenmodes into more frequencies.
\end{itemize}
It is hoped that the current approach will help to develop optimized schemes 
and selecting more appropiate ones taking into account different mesh geometries.
We also hope this analysis to be useful to design mesh stretching strategies that 
reduce dispersion errors.

On a further work, we plan to use this method to test more
differential schemes. Since non-linearity of the differential operator
is not a requirement of the developed methodology, the analysis will be
conducted on non-linear schemes.